\documentclass[a4paper]{article}

\usepackage{color}

 \usepackage{graphicx}

 \usepackage{amsmath,amsfonts,amssymb}

\usepackage{authblk}

\usepackage[boxed,ruled]{algorithm2e}
\usepackage{multirow}
\usepackage[section]{placeins}

\usepackage{float}

\usepackage{hyperref}

                {\par \noindent \textbf{Proof #1:} \ignorespaces}%
                {\hfill $\square$\\ \par}

 \newcommand{\R}{\mathbb{R}}
 
 \newcommand{\diag}{\mathrm{diag}}

 \makeatletter
\newcommand*\bigcdot{\mathpalette\bigcdot@{.5}}
\newcommand*\bigcdot@[2]{\mathbin{\vcenter{\hbox{\scalebox{#2}{$\m@th#1\bullet$}}}}}
\makeatother

\newcommand{\ddt}[1]{\frac{\mathrm{d} #1}{\mathrm{dt}}}
 \newcommand{\eps}{\varepsilon}
 \renewcommand{\H}{{\mathcal H}}

\newcommand{\ytrans}{y^\dagger}

\newcommand{\argmin}{\operatornamewithlimits{argmin}}

\begin{document}
\title{ \bf Nonparametric goodness-of-fit testing for parametric covariate models in pharmacometric analyses
}

\author{Niklas Hartung$^{1}$, Martin Wahl$^{2}$, Abhishake Rastogi$^{1}$, and
Wilhelm Huisinga$^{1,\ast}$}

\date{\relax}

\maketitle
\noindent
$^1$Institute of Mathematics, Universit\"at Potsdam, Germany\\[1ex]
$^2$Institute of Mathematics, Humboldt-Universit\"at zu Berlin, Germany\\[1ex]
$^\ast$corresponding author\\
Institute of Mathematics, Universit\"at Potsdam\\
Karl-Liebknecht-Str. 24-25, 14476 Potsdam/Golm, Germany\\
Tel.: +49-977-59 33, Email: huisinga@uni-potsdam.de 

\subsubsection*{Conflict of Interest/Disclosure}
The authors declare no conflict of interest.

\subsubsection*{Keywords}
Covariate modeling, maturation, nonparametric goodness-of-fit testing, statistical learning, Tikhonov regularization, reproducing kernel Hilbert spaces.

\subsubsection*{Acknowledgements}
This research has been partially funded by Deutsche Forschungsgemeinschaft (DFG) -- SFB1294/1 -- 318763901. 
Fruitful discussions with Markus Rei{\ss} (Humbold-Universit\"at zu Berlin, Germany), Gilles Blanchard (Universit\'e Paris-Saclay, France) and Matthias Holschneider (Universit\"at Potsdam, Germany) are kindly acknowledged.


\newpage

\begin{abstract}
The characterization of covariate effects on model parameters is a crucial step during pharmacokinetic/pharmacodynamic analyses.
While covariate selection criteria have been studied extensively, the choice of the functional relationship between covariates and parameters, however, has received much less attention. 
Often, a simple particular class of covariate-to-parameter relationships (linear, exponential, etc.) is chosen ad hoc or based on domain knowledge, and a statistical evaluation is limited to the comparison of a small number of such classes.
Goodness-of-fit testing against a nonparametric alternative provides a more rigorous approach to covariate model evaluation, but no such test has been proposed so far.
In this manuscript, we derive and evaluate nonparametric goodness-of-fit tests for parametric covariate models, the null hypothesis, against a kernelized Tikhonov regularized alternative, transferring concepts from statistical learning to the pharmacological setting.
The approach is evaluated in a simulation study on the estimation of the age-dependent maturation effect on the clearance of a monoclonal antibody. 
Scenarios of varying data sparsity and residual error are considered. 
The goodness-of-fit test correctly identified misspecified parametric models with high power for relevant scenarios.
The case study provides proof-of-concept of the feasibility of the proposed approach, which is envisioned to be beneficial for applications that lack well-founded covariate models. 
\end{abstract}

\newpage

\section*{Introduction}

Pharmacokinetic/pharmacodynamic (PK/PD) models are used to describe drug concentrations/effect over time in a group of patients under treatment. 
Covariate models are employed to describe the effect of patient characteristics (covariates) on model parameters, and play a crucial role in PK/PD model building. 
Common covariates are body weight, age, concentrations of important biomarkers (e.g., plasma creatinine) or genetic disposition (e.g., CYP450 polymorphisms). 
Additional random variability not explainable by covariates is accounted for through random effects \cite{standing2017}.
Typically, there is detailed knowledge on the model structure (often specified in terms of compartmental models), while much less is known on how covariates impact drug kinetics/effects via the model parameters.

Covariate selection criteria in pharmacometric analyses have been studied extensively \cite{jonsson1998,ribbing2007,hutmacher2015}.
In contrast, the choice of the functional relationship between covariates and parameters has received less attention, and often, a particular parametric class of covariate-to-parameter relationships (linear, exponential, etc.) is chosen \emph{ad hoc} \cite{joerger2012}.
To evaluate the appropriateness of a parametric covariate model class, a choice amongst a few candidate classes is sometimes made using likelihood ratio tests \cite{dartois2007} or information-theoretic criteria \cite{akaike1974,schwartz1978}. 
However, such a comparison depends on the classes considered, and furthermore does not reveal whether \emph{any} of these classes is compatible with the PK/PD data. 

Several methods have been proposed to overcome these limitations of a pre-specified parametric covariate-to-parameter relationship. 
In the so-called two-stage approach, a covariate model is obtained by first estimating individual parameters from individual patient data, then either choosing a suitable parametric class based on graphical analysis or solving a nonparametric regression problem \cite{dartois2007,mandema1992}. 
While this approach is feasible if parameters can be identified from individual data, it is not so in the more realistic scenario of sparse and noisy data, where the true covariate-to-parameter relationships may be masked or wrongly attributed to other parameters \cite{Savic09}. 
Covariate models using regression splines or neural networks have been proposed in \cite{Lai06}.
Using a small number of quadratic splines at fixed quantiles of the covariate distribution (or similarly, a small neural network), the authors were able to derive a more flexible covariate-to-parameter relationship than the commonly used classes of parametric models.
Another approach was presented in \cite{Mesnil98}, where a nonparametric maximum likelihood algorithm was developed and applied to clinical data. 
By estimating the joint distribution of parameters and covariates, it allowed to derive a nonparametric covariate-to-parameter relationship. 

While parametric as well as nonparametric approaches have been used for the estimation problem, a nonparametric goodness-of-fit test for parametric covariate models is still lacking.
For model evaluation and selection, however, a statistically sound comparison against a nonparametric alternative would be highly desirable
since it allows to challenge the functional form of the covariate-to-parameter relationship more critically.

In case of a direct covariate-to-\textit{observable} relationship (also called nonparametric regression, or a direct problem), goodness-of-fit testing has been extensively studied in the statistical literature \cite{haerdle1993,horowitz2001}, with applications in other fields such as econometrics \cite{hausman1978}.
A standard construction is based on a mean squared distance between a parametric and a nonparametric estimator. 
For instance, \cite{haerdle1993,horowitz2001} employ a Nadaraya-Watson estimator, but many other nonparametric estimators work as well. 
The direct problem, however, is not relevant to our context, since the structural model, i.e., the parameter-to-observable relationship, is based on a class of models with established trust in the pharmacometric community; 
in the case of PK models with further support from reducing more detailed physiologically-based pharmacokinetic models \cite{pilari2010}. 
Consequently, in our setting covariate models link to (unobservable) parameters rather than to direct observations. 
For the estimation of a covariate-to-\textit{parameter} model in a nonlinear parameter-to-observable relationship (also called nonlinear statistical inverse problem), a popular estimation method is Tikhonov regularization. 
Regularization methods have been studied in many different contexts, including inverse and ill-posed problems \cite{osullivan1990,engl1996} and, more recently, unsupervised learning \cite{caponnetto2007,smale2007,lu2013,rastogi2020}. 
In the context of statistical learning, the framework of reproducing kernel Hilbert spaces (RKHS) plays a central role to address both computational and theoretical questions \cite{aronszajn1950,cucker2002,schoelkopf2002,micchelli2005,alvarez2012}. 
In case of the direct problem, this framework is known under the name of multitask learning \cite{evgeniou2005}.

In this manuscript, we propose and evaluate nonparametric goodness-of-fit tests for parametric covariate models, transferring concepts developed in statistical learning to the pharmacological setting. 
We generalize known goodness-of-fit tests for the direct problem (e.g. based on kernel density estimation by \cite{haerdle1993}) to nonlinear statistical inverse problems and also present a tailored numerical approach for required computation of kernelized Tikhonov regularizers.
We then demonstrate proof-of-concept in a relevant pharmacological application, the estimation of an age effect on maturation of drug-metabolizing enzymes. 
As a first step, here, we focus on the case without random effects.

\section*{Methods}\label{sec:methods}

\subsection*{Nonparametric goodness-of-fit testing} \label{sec:gof-test}


We consider the statistical model
\begin{equation}\label{eq:nlinvprob}
y_i = G(\theta_i,x_i) + \varepsilon_i, \quad \theta_i = f(x_i), \quad i\in\{1,\ldots,n\},
\end{equation}
with i.i.d.~observations $\big(x_i,y_i\big)_{i=1}^n$ consisting of \emph{covariates} $x_i\in\mathcal{X}\subset\R^{n_x}$ and (noisy) \emph{observations} $y_i\in\R^q$, unobserved \emph{parameters} $\theta_i\in\Theta\subset \R^p$, a nonlinear function $G:\Theta\times\mathcal{X}\rightarrow \R^q$ called the \emph{mechanistic model}, independent centered noise $\eps_i$, and a \emph{covariate-to-parameter mapping} $f:\mathcal{X}\rightarrow\Theta$.

We assume the mechanistic model $G$ to be known, which might represent a component of the solution of a  system of ordinary differential equations (ODEs), observed at different time points $t_1,...,t_q$, or a transformation thereof. 
The direct dependency of $G$ on $x_i$ allows to model individualized doses, and to model only particular aspects of a covariate model; this will also be used in our simulation study.
Even for the simplest model, $G$ depends nonlinearly on the parameters $\theta$.
The covariate model $f$ is assumed to be unknown.
We assume that a particular parametric class of functions $\big\{f_\tau, \tau\in\mathcal{T}\subset\R^{n_\tau}\big\}$ is given, chosen \emph{ad hoc} or from domain knowledge, and aim to  evaluate the hypothesis that the covariate model $f$ belongs to this parametric class. Thus, we consider the test problem
\begin{equation}\label{eq:test-problem}
H_0: f\in \{f_\tau, \tau\in\mathcal{T}\}\quad \text{vs.}\quad H_1: f\not\in \{f_\tau, \tau\in\mathcal{T}\}.
\end{equation}
The precise formulation of the considered alternative, i.e., the space of functions for $H_1$, and the test statistics are still to be specified.

\paragraph{Testing against a nonparametric alternative.}

First, we consider a \emph{nonparametric} alternative of the form
$f\in \H\setminus \{f_\tau, \tau\in\mathcal{T}\}$,
with a suitably chosen vector-valued RKHS $\H$ of functions $h:\mathcal{X}\rightarrow \mathbb{R}^p$.
Briefly, an RKHS is specified via a kernel function $k$ that is used to define a basis for $\H$ (see also \eqref{eq:shortdual}).
Background and references on vector-valued RKHS are stated in Section~\ref{sec:formulations}.

We start by defining two natural estimators under the null and the alternative, namely the least squares estimator
\begin{equation}\label{eq:least-squares}
f_{\hat \tau}\quad\text{with}\quad\hat \tau := \argmin_{\tau\in\mathcal{T}}\left[\sum_{i=1}^n \Big\|y_i - G\big(f_\tau(x_i),x_i\big)\Big\|^2\right]
\end{equation}
and the Tikhonov regularized estimator (see e.g. \cite{caponnetto2007,rastogi2020})
\begin{equation}\label{eq:tikhonov-nonpar}
\hat f^{(\lambda)} := \argmin_{h\in\H}\left[\sum_{i=1}^n \Big\|y_i - G\big(h(x_i),x_i\big)\Big\|^2 + \lambda \big\|h\big\|^2_{\H}\right]
\end{equation}
with regularization parameter $\lambda>0$, respectively\footnote{$\|\cdot\|$ and $\|\cdot\|_\H$ denote the Euclidean and the RKHS norm, respectively, for the latter see also Section~\ref{sec:formulations}.}. Based on these estimators, we consider the test statistic $T_1=T_1^{(\lambda)}$, defined as
\begin{equation}\label{eq:T1}
T_1 := \sum_{i=1}^n \Big\|G\Big(f_{\hat \tau}(x_i),x_i\Big) - G\Big(\hat f^{(\lambda)}(x_i),x_i\Big)\Big\|^2.
\end{equation}
The statistic evaluates the estimated covariate-to-parameter relationships in the \textit{space of observations}, i.e., after mapping of $G$; alternatively, a test statistic could be directly based on the difference of the estimated covariate-to-parameter relationships in the \textit{space of parameters}. 
In \cite{haerdle1993}, also smoothed versions of the parametric estimate $f_{\hat\tau}$ were considered. 
Both variations are introduced in Section~\ref{sec:goftest-extra} and all considered statistics are compared in the \nameref{sec:discussion}. 

\paragraph{Testing against a combined parametric-nonparametric alternative.}

In addition to testing against a nonparametric alternative, 
we consider a \emph{combined parametric-nonparametric alternative} of the form
$f = f_\tau + h$ with $(\tau, h)\in\mathcal{T}\times\H$, but $f\not\in\{f_\tau: \tau\in\mathcal{T}\}$. 
If $f_\tau \in \H$ for each $\tau\in\mathcal{T}$, then the two classes of functions considered under the nonparametric and combined parametric/nonparametric alternatives are identical. 
The reason we introduce this reformulation is to consider a different test statistic which uses the Tikhonov-type regularization scheme
\begin{equation}
\label{eq:tikhonov-combined}
\tilde f^{(\lambda)} := \argmin\limits_{\substack{f=f_\tau+h,\\ (\tau,h)\in\mathcal{T}\times\H}}\left[\sum_{i=1}^n\Big\|y_i - G\big(f_\tau(x_i) + h(x_i),x_i\big)\Big\|^2 + \lambda \|h\|_{\H}^2\right],
\end{equation}
with regularization parameter $\lambda>0$. 
Based on this, we define the second test statistic
\begin{equation}\label{eq:T3}
T_2 := \sum_{i=1}^n \Big\|G\Big(f_{\hat \tau}(x_i),x_i\Big) - G\Big(\tilde f^{(\lambda)}(x_i),x_i\Big)\Big\|^2,
\end{equation}
which is analogously defined as $T_1$ in the nonparametric case above.

The combined model $f = f_\tau+h$ is appealing from a modeling point of view because it can be used to penalize deviations from the parametric covariate model only, but not the parametric covariate model itself. For example, for large values of $\lambda$, the purely nonparametric covariate model $\hat f^{(\lambda)}$ shrinks to 0, where the mechanistic model $G$ might even be undefined. In contrast, the combined model $\tilde f^{(\lambda)}$ shrinks towards the parametric part and hence, does not suffer from the undesired behavior of the purely nonparametric model.

\paragraph{Critical values of test statistics.} 

Critical values $c_\alpha$ for the level $0<\alpha<1$ and both test statistics (generically denoted $T$ here) are approximated by a Monte Carlo procedure. Based on the data $(x_i,y_i)_{i=1}^n$, the least squares estimator $\hat\tau$ for the parametric null model is determined. Then, $M$ synthetic datasets 
$(x_i,y^{(1)}_i)_{i=1}^n,\ldots,(x_i,y^{(M)}_i)_{i=1}^n$
are simulated under the approximate null model (with $\hat\tau$ instead of the unknown true value $\tau^*$), 
i.e. for $m=1,\ldots,M$, 
\[y^{(m)}_i = G\big(f_{\hat\tau}(x_i),x_i\big) + \varepsilon_i^{(m)},\quad i=1,\ldots,n,\] 
where $\varepsilon_i^{(m)}$ are i.i.d.~realizations of the noise (we assume for simplicity that the distribution of residual errors is known).
Based on the synthetic data, the statistic $T^{(m)}$ is computed as described previously. 
Then, letting $\hat F_T$ denote the empirical distribution of $T$, we choose $c_\alpha = \hat F_T^{-1}(\alpha)$ as critical value for the test (since large values of $T$ favour the alternative for all considered statistics).

\subsection*{Efficient algorithms for estimation in a nonparametric model} \label{sec:numerics}

The calculation of the test statistics in \eqref{eq:T1} and \eqref{eq:T3} requires different optimization problems to be solved, namely the least squares problem \eqref{eq:least-squares} to determine $f_{\hat \tau}$, the Tikhonov regularization problem \eqref{eq:tikhonov-nonpar} to determine $\hat f^{(\lambda)}$, and the combined least squares/Tikhonov regularization problem \eqref{eq:tikhonov-combined} to determine $\tilde f^{(\lambda)}$.
Whereas the parametric problem is usually low-dimensional, the others are high-dimensional. 
Since the mechanistic model $G$ is assumed to depend nonlinearly on the parameters $\theta$, none of these problems can be solved in closed form. 
As will be shown in Section~``\nameref{sec:benchmark}'', general-purpose optimizers perform poorly on the high-dimensional nonlinear problems and are very sensitive to the choice of the initial conditions, motivating the need for more tailored numerical approaches.

\paragraph{Parametrization of RKHS problems.}

The representer theorem in the theory of RKHS guarantees the existence of $f^{(\lambda)}$ within the finite-dimensional space
\begin{equation}\label{eq:shortdual}
\left\{h_\alpha \in \H \Big|h_\alpha:=\sum_{i=1}^n k(\cdot,x_i)\alpha_i\; \text{ with } \; \alpha_1,\ldots,\alpha_n\in \R^p\right\},
\end{equation}
where $k$ is the kernel associated to the RKHS $\H$ \cite{micchelli2005,alvarez2012}.
A finite-dimensional formulation is a prerequisite for solving \eqref{eq:tikhonov-nonpar} numerically. 
Besides the so-called \emph{dual formulation} \eqref{eq:shortdual} of the finite-dimensional optimization problem, two other finite-dimensional formulations can be obtained for special types of kernels (admitting finite-dimensional feature map representations). 
In this case, the dimension $np$ of the dual formulation can be reduced further through a reparametrization (leading to the so-called \emph{primal} and \emph{mixed} formulations). 
This technique is described in detail in Section~\ref{sec:formulations} and exploited in the considered simulation study.
For ease of readability, in the main text we always refer to the function $\hat f^{(\lambda)}$ solving optimization problem \eqref{eq:tikhonov-nonpar} without specifying its underlying parametrization.

\paragraph{Estimation algorithms.}

To solve the least squares problem and obtain the parametric estimate $f_{\hat \tau}$, we use the Levenberg-Marquardt (LM) algorithm, a robust gradient-based method for solving nonlinear least squares problems \cite{levenberg1944,marquardt1963}.

To solve the Tikhonov regularization problem \eqref{eq:tikhonov-nonpar}, we propose a three-step algorithm that first solves easier approximate problems, and uses the solution of each step to obtain improved initial guesses for the subsequent step (see pseudocode in Algorithm~\ref{algo:nonpar}): 
\begin{enumerate}
\item \emph{ParDir:} determine a \underline{par}ametric estimate $f_{\hat \tau}$ via LM algorithm, then solve the \underline{dir}ect nonparametric (RKHS) problem analytically (see Section~\ref{sec:lininvprob}) by considering $f_{\hat \tau}(x_i)$ as surrogate for the unobservable parameters;
\item \emph{AlyLin:} \underline{a}na\underline{ly}tically solve a sequence of \underline{lin}earized nonparametric problems (see Section~\ref{sec:lininvprob} for derivation);
\item \emph{Nonlin:} finally, solve the original \underline{nonlin}ear nonparametric problem \eqref{eq:tikhonov-nonpar} employing a quasi-Newton method.
\end{enumerate}
In this way, local minima of \eqref{eq:tikhonov-nonpar} can be avoided more successfully.
An analysis and benchmark of this algorithm against several alternative approaches is shown for the simulation study, where it clearly outperforms general-purpose optimizers in terms of robustness and runtime.
\begin{algorithm}[h]
\DontPrintSemicolon
\SetKwFunction{LevenbergMarquardt}{LevenbergMarquardt}
\SetKwFunction{LinearizeModel}{LinearizeModel}
\SetKwFunction{QuasiNewton}{QuasiNewton}
\SetKwData{niter}{niter}
\BlankLine
\BlankLine
\tcp{Step 1:~"ParDir"}
$\text{res} := \left[\tau \,\mapsto\, \Big(G(f_\tau(x_1),x_1)-y_1,\ldots,G(f_\tau(x_n),x_n)-y_n\Big)\right]$\;
$\hat\tau \longleftarrow \text{minimize } \|\text{res}(\tau)\|^2 \text{ using Levenberg-Marquart with initial guess } \tau_0$\;
$Q_\text{dir} :=\left[h\mapsto\sum\limits_{i=1}^n \|f_{\hat\tau}(x_i)-h(x_i)\|^2 + \lambda \|h\|^2_\H\right]$\;
$h_\text{dir} \longleftarrow \text{minimize }Q_\text{dir}(h) \text{ over } \H \text{ analytically }$\;
\BlankLine
\BlankLine
\tcp{Step 2:~"AlyLin"}
$h_\text{lin}^{(0)} \longleftarrow h_\text{dir}$\;
\For{$s =  1:\niter$}{
$G^{(s)}_\text{lin} \longleftarrow \text{linearize model } G \text{ at function } h_\text{lin}^{(s-1)}$\;
$Q^{(s)}_\text{lin} := \left[h \mapsto \sum\limits_{i=1}^n \Big\|y_i-G^{(s)}_\text{lin}\big(h(x_i),x_i\big)\Big\|^2 + \lambda \big\|h\big\|^2_\H\right]$\;
$h_\text{lin}^{(s)} \longleftarrow \text{ minimize } Q^{(s)}_\text{lin}(h) \text{ over } \H \text{ analytically }$
}
\BlankLine
\BlankLine
\BlankLine
\tcp{Step 3:~"Nonlin"}
$Q := \left[h \mapsto \sum\limits_{i=1}^n \Big\|y_i-G\big(h(x_i),x_i\big)\Big\|^2 + \lambda \big\|h\big\|^2_\H\right]$\;
$h_\text{nonlin} \longleftarrow \text{minimize } Q(h) \text{ using quasi-Newton with initial guess } h_\text{lin}^{(\niter)}$\;
\BlankLine
\BlankLine
\BlankLine
\tcp{Output}
$\hat f^{(\lambda)} \longleftarrow h_\text{nonlin}$\;
\KwRet{$\hat f^{(\lambda)}$}
\caption{\label{algo:nonpar}
ParDir-AlyLin-Nonlin for problem \eqref{eq:tikhonov-nonpar}
} 
\end{algorithm}

Finally, a variant of Algorithm~\ref{algo:nonpar} can be used to obtain the combined parametric/nonparametric estimate $\tilde f^{(\lambda)}$ solving \eqref{eq:tikhonov-combined}. 
The pseudocode for this Algorithm~\ref{algo:combined} is provided in Section~\ref{sec:app-algo2}.

\paragraph{Implementation.}

The proposed estimation algorithms and goodness-of-fit tests have been implemented in \textbf{R} version~3.5.1 \cite{R_base}. For matrix algebra, \textbf{R} package \texttt{Matrix} version~1.2-14 was used \cite{R_Matrix}. The Levenberg-Marquardt algorithm was taken from \textbf{R} package \texttt{minpack.lm} version~1.2-1, an interface to the Fortran library MINPACK \cite{R_minpack}. The general-purpose optimizers implemented in base \textbf{R} function \texttt{optim} were used for the quasi-Newton method (BFGS algorithm) and simulated annealing.
The code used during the analysis is publicly available at \url{https://zenodo.org/record/4273796}.

\subsection*{Simulation study: Effect of enzyme maturation on drug clearance}\label{sec:setup}

\paragraph{Context of the simulation study setup.}

A functional relationship between body weight and drug disposition parameters, called allometric scaling, is well established in the pharmacometric literature, see e.g.~\cite{anderson2009}. 
In young children (in particular neonates and infants), however, this weight-effect is not sufficient to describe pharmacokinetic data and an additional weight-independent impact of (young) age on drug clearance is accounted for by a \emph{maturation function} \cite{robbie2012,holford2013}.
Many different parametric maturation functions have been proposed in the literature (see e.g.~\cite{germovsek2017} for an overview).
Therefore, goodness-of-fit testing for parametric covariate models is of particular importance in this context. 

The setup of our simulation study ``Effect of enzyme maturation on drug clearance'' was motivated by the meta-analysis in \cite{robbie2012}, which estimated the maturation effect of the monoclonal antibody palivizumab against the respiratory syncytial virus infections in young children.
We translated their setting to our statistical framework as follows:
\paragraph{Covariates.} In \cite{robbie2012}, the covariates (post-gestational) age, weight, gender, ethnicity, and presence/absence of chronic lung disease are considered. 
For ease of presentation, we focussed on the covariates (post-natal) age $a$ in [years] and body weight $w$ in [kg], i.e.~$x=(a,w)$. 
We assumed a uniform age distribution between 0 and 20 years and an age-dependent body weight distribution according to an empirical model by Sumpter et al.~\cite{sumpter2011}. 

\paragraph{Mechanistic model $G$.} As in \cite{robbie2012}, we assumed a two-compartment PK model 
\begin{align}
V_1\, \ddt{C_1} &=  Q(C_2-C_1) - \text{CL} \cdot C_1,\label{eq:ODE-1}\\
V_2\, \ddt{C_2} &=  Q(C_1-C_2)\label{eq:ODE-2},
\end{align}
with drug concentrations $C_1$, $C_2$ in [mg/L] in the central and peripheral compartments with volumes $V_1$, $V_2$ in [L], respectively, inter-compartmental flow $Q$ in [L/day] and clearance CL in [L/day]. 
The initial conditions were $C_1(0) = \text{D}_w \cdot w/{V_1}$ and $C_2(0) = 0$ with an i.v.\ bolus administration of $\text{D}_w = 15$ [mg/kg body weight]. 
We also considered a multiple dosing scenario with 30-day dosing intervals.
The model can be solved analytically, see Section~\ref{sec:Gmaturation}.

The model is parametrized in terms of the parameters $\theta=(\text{CL}, V_{1}, Q, V_{2})$. In \cite{robbie2012}, the covariate-to-parameter relationship comprises a maturation part (depending on age $a$) and an allometric part (depending on weight $w$): 
\begin{align*}
\text{CL} &= \text{CL}_0 \cdot\text{mat}(a) \biggl(\frac{w}{w_\text{ref}}\biggr)^{\frac34}; &&&
\text{V}_1 &= \text{V}^*_{1}  \biggl(\frac{w}{w_\text{ref}}\biggr);\\
\text{Q} &= \text{Q}^* \biggl(\frac{w}{w_\text{ref}}\biggr)^{\frac34}; &&&
\text{V}_2 &= \text{V}^*_{2}  \biggl(\frac{w}{w_\text{ref}}\biggr);
\end{align*}
with reference body weight $w_\text{ref}= 70~\text{kg}$.
Since allometric scaling in the stated form is widely accepted, we considered it as part of the ODEs defining the mechanistic model $G$, 
\begin{align} 
\text{V}^*_{1}  \biggl(\frac{w}{w_\text{ref}}\biggr)\, \ddt{C_1} &=  \text{Q}^* \biggl(\frac{w}{w_\text{ref}}\biggr)^{\frac34}(C_2-C_1) - \text{CL}^* \biggl(\frac{w}{w_\text{ref}}\biggr)^{\frac34} \, C_1, \label{eq:ODEs-including-allo-scaling-eq-1}\\
\text{V}^*_{2}  \biggl(\frac{w}{w_\text{ref}}\biggr)\, \ddt{C_2} &=  \text{Q}^* \biggl(\frac{w}{w_\text{ref}}\biggr)^{\frac34}(C_1-C_2). \label{eq:ODEs-including-allo-scaling-eq-2}
\end{align}
and considered the weight-normalized parameters $\theta = (\text{CL}^*, V^*_{1}, Q^*, V^*_{2})$ as unknown.
The observed quantity was 
\[G(\theta,x) = \big(\ln C_1(t_1),\ldots,\ln C_1(t_q)\big)\] 
at fixed time points $t_1,\ldots,t_q$. Finally, we assumed normally distributed additive noise, i.e. $(\varepsilon_i)\sim_\text{iid}\mathcal{N}(0,\sigma^2)$ with $\sigma>0$ known. 

\paragraph{Covariate-to-parameter relationship.} 
To generate the virtual clinical data, a saturable exponential maturation function as in \cite{robbie2012}
\[\text{CL}^*(a) = (1- \alpha e^{-\beta\cdot a})\,\text{CL}^*_\text{max}\]
and the resulting covariate model
\begin{equation*}
f_\tau(a) = \Big(\,\text{CL}^*(a), \, V^*_{1}, \,  Q^*, \,V^*_{2}\Big)
\end{equation*}
were used, with parameter values listed in Tab.~\ref{tab:parameters}.

\begin{table}[htp]
\centering
\begin{tabular}{r|rl}
Parameter  & Value  & [unit]\\
\hline
$\alpha$ & 0.589$^\dagger$ & --\\
$\beta$ & 0.133 & [1/year]\\
$\text{CL}^*_\text{max}$ & 198 & [mL/day]\\
$V^*_{1}$ & 4090 & [mL]\\
$Q^*$ & 879 & [mL/day]\\
$V^*_{2}$ & 2230 & [mL]\\
\end{tabular}
\caption{\label{tab:parameters} {\it 
Parameters of the covariate model used for simulation, taken from \cite{robbie2012}.
$\dagger$: In the original publication, $\alpha = 0.411$ was reported, inconsistent with the remaining results shown in the article. In contrast, $0.589 = 1 - 0.411$ was consistent with all other results, hence we used this value for the simulation study.
}}
\end{table}

\begin{figure}[htp!]
\centering
\includegraphics[width=\textwidth]{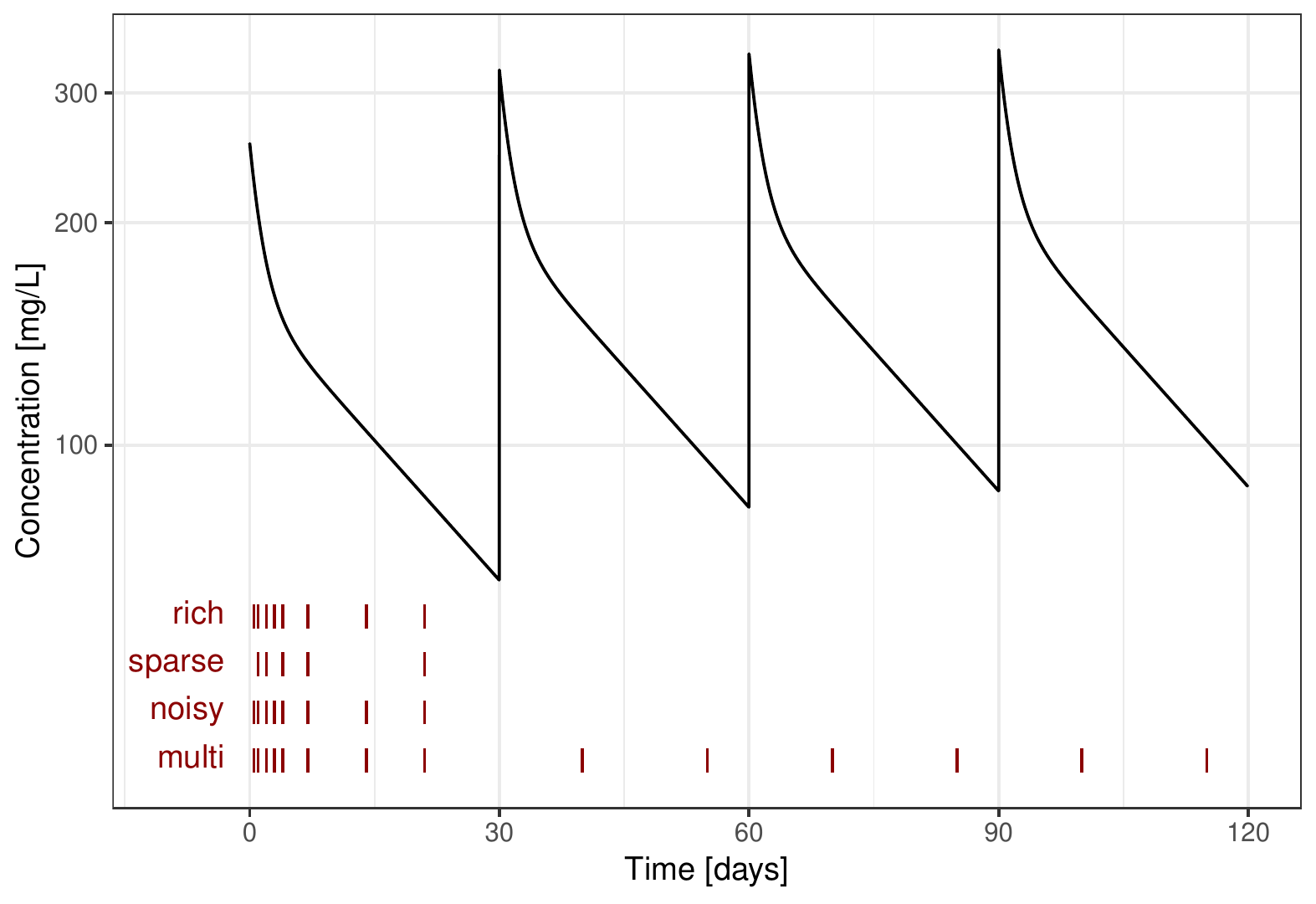}
\caption{\label{fig:pk} {\it
Typical plasma concentration-time profile for a reference adult (70~kg body weight) in the simulation study ``Effect of enzyme maturation on drug clearance'', based on PK parameters for Palivizumab from \cite{robbie2012}. 
Four 30-day dosing cycles with a dose of 15~mg/kg body weight are simulated.
The sampling times of the four considered scenarios (rich, sparse, noisy, multi) are indicated in red.
}}
\end{figure}

\paragraph{Simulation scenarios.} 
We considered four simulation scenarios, varying in number of individuals, noise level, and sampling times (see Tab.~\ref{tab:scenarios}). 
A typical prediction with the mechanistic model is shown in Fig.~\ref{fig:pk}, along with the sampling times of the four considered scenarios. 
Three scenarios (rich, sparse, noisy) contained sampling points from a single dosing interval, while scenario ``multi'' had sampling times in four dosing intervals. 
Since the clearance is mainly informed by the terminal phase of a dosing cycle, the multiple dosing scenario is expected to allow for more precise estimates compared to the corresponding single dose scenario (noisy). 
The noise level reported in \cite{robbie2012} was $\sigma\approx0.24$, hence scenarios ``rich'' and ``sparse'' were less noisy, while scenarios ``noisy'' and ``multi'' were more noisy than the original model. 
Exemplary simulated data for each of the four scenarios are shown in Section~\ref{sec:sim-data}.

\noindent
\begin{table}[htp!]	
\centering
\begin{tabular}{r||r|c|l}
Scenario & $n$~ & $\sigma$ & Observation timepoints [days]\\
\hline
rich    & 100 & 0.1 & 0.5, 1, 2, 3, 4, 7, 14, 21\\
sparse  & 20  & 0.1 & ~~~~~\,1, 2, ~~~\,4, 7, ~~~~~21\\
noisy   & 100 & 0.3 & 0.5, 1, 2, 3, 4, 7, 14, 21\\
multi   & 100 & 0.3 & 0.5, 1, 2, 3, 4, 7, 14, 21, 40,  55,  70,  85, 100, 115\\
\end{tabular}
\caption{\label{tab:scenarios} {\it
Four scenarios considered in the simulation study ``Effect of enzyme maturation on drug clearance'', differing in number of individuals $n$, standard deviation $\sigma$ of the residual error, and observation time points.
}}
\end{table}
\vskip 5mm

\FloatBarrier

\paragraph{Parametric covariate model classes for goodness-of-fit testing.}

Three different classes of covariate-to-parameter relationships were considered that differed in the parametrization of the $\text{CL}^*$ function.
All three classes have been reported in the literature, see \cite{germovsek2017}.
\begin{itemize}
\item Class based on \emph{saturable exponential} $\text{CL}^*$ functions
\[f_\tau(a) = \Big( (1- \alpha e^{-\beta\cdot a})\,\text{CL}^*_\text{max}, \, V^*_{1}, \,  Q^*, \,V^*_{2}\Big),\]
with $\tau = (\alpha,\,\beta,\,\text{CL}^*_\text{max}, \, V^*_{1}, \,  Q^*, \,V^*_{2})$. This parametric class allowed us to evaluate the Type~I error of the goodness-of-fit tests, since it contains the model used to generate the virtual data.

\item Class based on \emph{affine linear} $\text{CL}^*$ functions 
\[f_\tau(a) = \Big( \alpha + \beta\, a, \, V^*_{1}, \,  Q^*, \,V^*_{2}\Big),\]
with $\tau = (\alpha,\,\beta, \, V^*_{1}, \,  Q^*, \,V^*_{2})$. 

\item Class based on \emph{Michaelis-Menten} type $\text{CL}^*$ functions
\[f_\tau(a) = \Big( \frac{\text{CL}^*_\text{max}\, a}{K_M+ a}, \, V^*_{1}, \,  Q^*, \,V^*_{2}\Big),\]
with $\tau = (\text{CL}^*_\text{max},\,K_M, \, V^*_{1}, \,  Q^*, \,V^*_{2})$. 
\end{itemize}

\paragraph{Choice of kernels and regularization parameters.}  
As stated in \eqref{eq:shortdual}, any nonparametric estimate of the covariate-to-parameter function will be of the form
\begin{equation} \label{eq:f-weighted-sum-of-kernels}
f = \sum_{i=1}^n k(\cdot,x_i)\alpha_i; \qquad  \alpha_1,...,\alpha_n\in \R^p.
\end{equation}
Since $f(x)=\theta$, the $l$-th entry of the parameter vector $\theta= (\text{CL}^*, V^*_{1}, Q^*, V^*_{2})$ corresponds to the $l$-th row of the kernel $k(\cdot,\cdot)$. 
For the RKHS $\H$ in the nonparametric alternative, an age-dependent diagonal kernel of the form
\begin{equation}\label{eq:kernel-nonpar}
k(a,a') = \begin{pmatrix}
\exp\left(-\frac{(a-a')^2}{2\text{b}^2}\right) & 0 & 0 & 0\\
0 & 1 & 0 & 0\\
0 & 0 & 1 & 0\\
0 & 0 & 0 & 1\\
\end{pmatrix}
\end{equation}
was assumed (allometric scaling, and hence the dependency on weight $w$, was modelled as part of the mechanistic model $G$). 
The first component (for parameter CL$^*$) is a Gaussian kernel with bandwidth parameter $b>0$, the $1$'s on the diagonal correspond to constant kernels. 
Using this kernel structure, the dependency of CL$^*$ on age was modelled as a weighted sum of Gaussians (see eq.~\eqref{eq:f-weighted-sum-of-kernels} above), whereas the other three parameters were constant (age-independent), since allometric scaling was part of the ODE model; see eqs.~\eqref{eq:ODEs-including-allo-scaling-eq-1}+\eqref{eq:ODEs-including-allo-scaling-eq-2}. 

For the combined parametric/RKHS model, a slightly different kernel was chosen because the age-independent components were already contained in the parametric part, leading to the kernel
\begin{equation}\label{eq:kernel-combined}
k(a,a') = \begin{pmatrix}
\exp\left(-\frac{(a-a')^2}{2\text{b}^2}\right) & 0 & 0 & 0\\
0 & 0 & 0  & 0\\
0 & 0 & 0 & 0\\
0 & 0 & 0 & 0\\
\end{pmatrix}.
\end{equation}
We chose $b=100$~[weeks] ($\approx$ 2 years) as bandwidth, representing 10\% of the simulated age range. The regularization parameters $\lambda>0$ in the Tikhonov regularization problems \eqref{eq:tikhonov-nonpar} and \eqref{eq:tikhonov-combined} were chosen by 5-fold cross-validation, striking a balance between goodness-of-fit (small $\lambda$) and generalizability to new data (larger $\lambda$). 

\section*{Results}\label{sec:results}

\subsection*{Estimated maturation functions based on cross-validated regularization parameters}
\label{sec:cv}

As a first step towards a nonparametric estimation of the maturation function, the regularization parameter $\lambda$ was estimated by 5-fold cross-validation.
We simulated one dataset per scenario (rich, sparse, noisy, multi) and determined the cross-validated estimate $\hat\lambda$ for the nonparametric estimator \eqref{eq:tikhonov-nonpar} and for the combined parametric/nonparametric estimator \eqref{eq:tikhonov-combined} for each of the three parametric classes. 
The estimated $\hat\lambda$ values were not sensitive to differences in the data scenarios. Of note, in the combined parametric/nonparametric model, $\hat\lambda$ strongly depended on the chosen parametric class, increasing from Michaelis-Menten, to affine linear, to the saturable exponential class.

\begin{figure}[htp!]
\centering
\includegraphics[width=\textwidth]{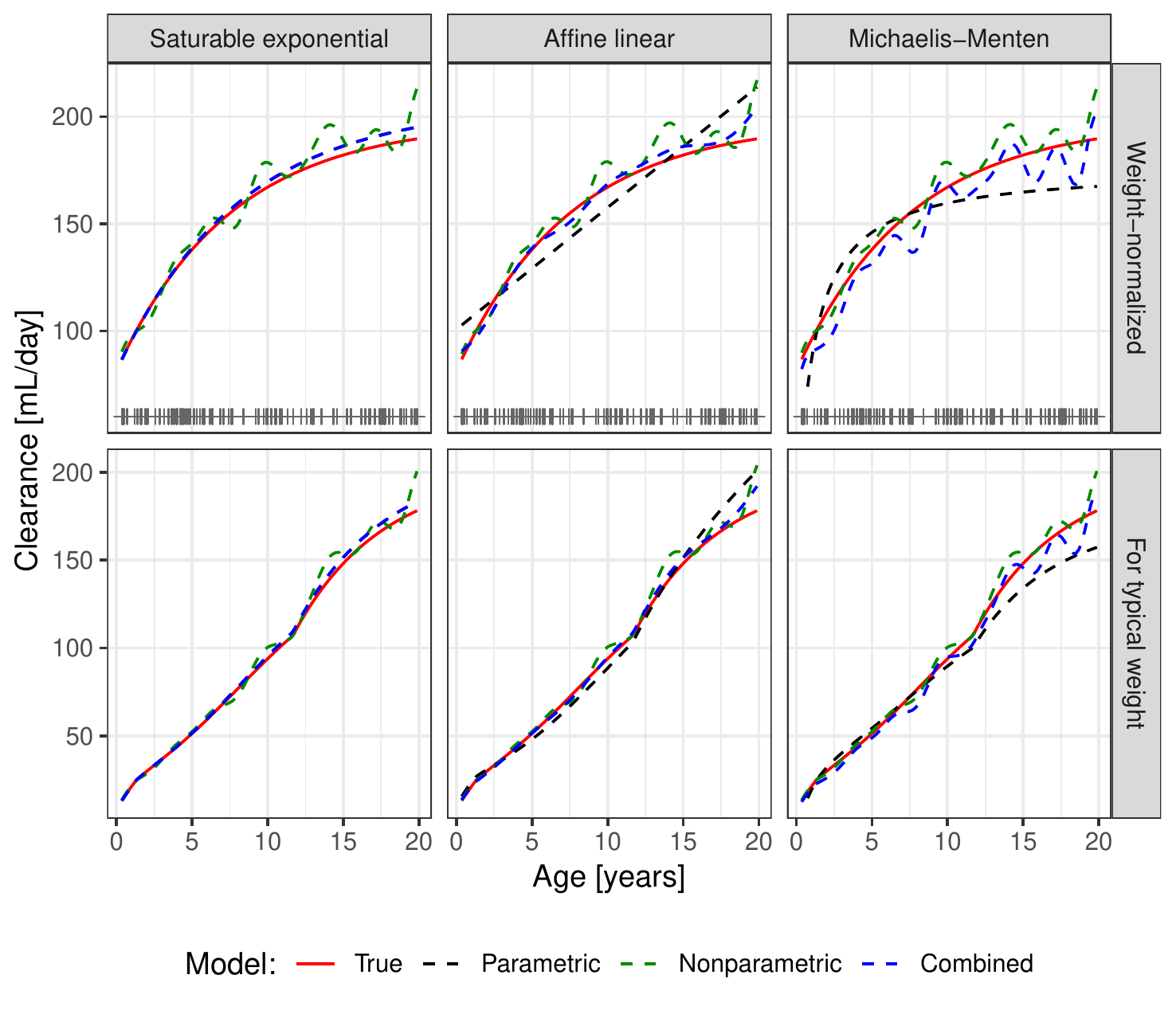}
\caption{\label{fig:matfun} {\it
Relationship between age $a$ and clearance predicted by parametric (black dashed line), nonparametric (green dashed line) and combined parametric/nonparametric (blue dashed line) estimates for one simulated dataset in scenario ``rich''. 
Top panel: weight-normalized clearance $\text{CL}^*(a)$ ; bottom panel: clearance $\text{CL}(a,w_\text{typ}(a))$ for the typical weight $w_\text{typ}(a)$ at a certain age $a$ (median weight predicted with the model by \cite{sumpter2011}). Each column corresponds to a different parametric class, left: saturable exponential (containing the true model, red solid line), middle: affine linear, and right: Michaelis-Menten. 
Grey crosses in top panel: age values in the simulated dataset.
}}
\end{figure}

See Fig.~\ref{fig:matfun} for an illustration of the estimated covariate-to-parameter relationships for the data-rich scenario. Due to the inverse problem character of the estimation problem (which depends on the sensitivity of $G$ to the parameters $\theta$), oscillations appear in the nonparametric and combined parametric/nonparametric estimates (see also Section~\nameref{sec:discussion}).
The effect of the larger penalization parameter in the saturable exponential class, in particular compared to the Michaelis-Menten class, was visible through a much smoother combined estimate $\tilde f^{(\hat\lambda)}$.

\subsection*{Efficient nonparametric estimation of the maturation function}
\label{sec:benchmark}
From a computational point of view, the proposed goodness-of-fit tests depend crucially on the performance of the numerical algorithms used to determine the observed test statistics and, through the Monte Carlo procedure, the critical values of the test statistics.
We therefore simulated 25 independent datasets for each of the four considered scenarios (rich, sparse, noisy, multi) for an evaluation in terms of convergence and runtime.
The mixed RKHS formulation (see Section~\ref{sec:formulations}) allowed to substantially reduce the dimensionality of the corresponding estimation problems, from $4n$ to $n+3$ parameters for the kernel structure \eqref{eq:kernel-nonpar} used in \eqref{eq:tikhonov-nonpar} and from $4n$ to $n$ for the kernel structure \eqref{eq:kernel-combined} used in \eqref{eq:tikhonov-combined}.
For ease of presentation, in this section we concentrate on Algorithm~\ref{algo:nonpar} in the nonparametric estimation problem \eqref{eq:tikhonov-nonpar}; a benchmark for Algorithm~\ref{algo:combined} in the combined parametric/nonparametric estimation problem \eqref{eq:tikhonov-combined} is provided in Section~\ref{sec:benchmark-combined}.

We benchmarked our proposed Algorithm~\ref{algo:nonpar} \textit{ParDir-AlyLin-Nonlin} against two commonly used general-purpose optimizers, namely \textit{quasi-Newton} (gradient-based) and \textit{simulated annealing} (gradient-free). In addition, we included two variants of Algorithm~\ref{algo:nonpar} that allow to further elucidate the impact of the different steps in Algorithm~\ref{algo:nonpar}: \textit{ParDir-Nonlin} (only steps 1 and 3) and \textit{ParDir-AlyLin} (only steps 1 and 2). 
The general-purpose optimizers were initialized with lognormally distributed initial conditions around 1, yielding a reasonable initial guess in the absence of more detailed knowledge. 
For Algorithm~\ref{algo:nonpar} and its variants, the parametric class of affine linear models was chosen in step 1 (see Fig.~\ref{fig:matfun}, middle panel), and the initial guess for its coefficients was lognormally distributed and having a plausible order of magnitude.
All lognormal distributions had a large coefficient of variation of $\approx 130\%$ (a log-variance of 1).

\begin{figure}[htp!]
\centering
\includegraphics[width=\textwidth]{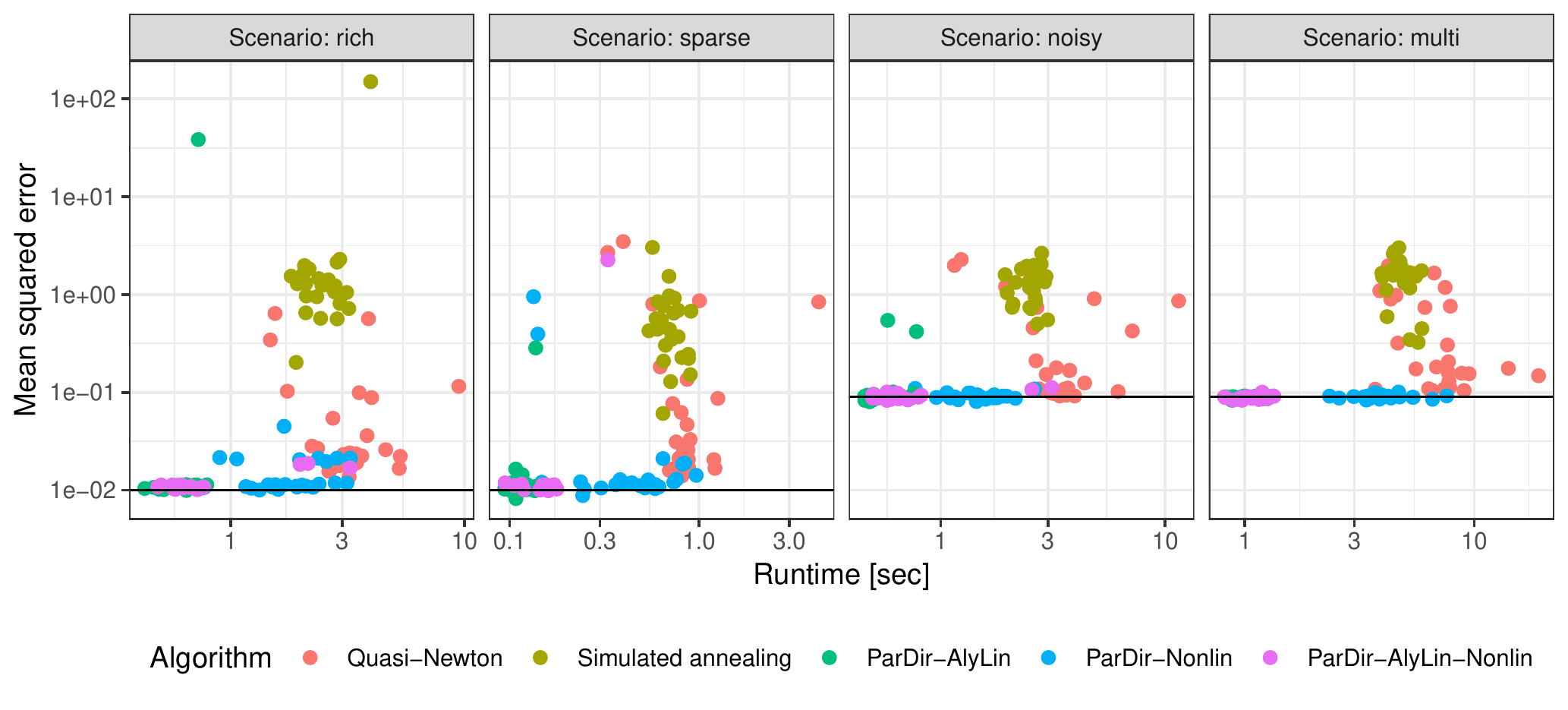}
\caption{\label{fig:benchmark} {\it
Benchmark of estimation algorithms for solving the Tikhonov regularization problem \eqref{eq:tikhonov-nonpar} in the simulation study ``Effect of enzyme maturation on drug clearance''. 
We simulated 25 independent datasets for each of the four considered data scenarios (rich, sparse, noisy, multi) and benchmarked the following estimation algorithms: a quasi-Newton method, simulated annealing, the proposed Algorithm~\ref{algo:nonpar} (steps 1-3); and two variants of it:  ParDir-AlyLin (only steps 1 and 2) and ParDir-Nonlin (only steps 1 and 3).
For the general-purpose optimizers, random positive initial conditions were chosen.
For optimizers starting with a parametric step, the parametric class of affine linear models was chosen, and the initial guess for its coefficients was lognormally distributed around values $\tau_0$ having the correct order of magnitude, with a coefficient of variation of $\approx 130\%$. 
Each dot represents runtime and mean squared error for one of the 25 simulated datasets.
The black horizontal line indicates the error variance $\sigma^2$ used to generate the data.
}}
\end{figure}

The benchmark results based on our simulation study are displayed in Fig.~\ref{fig:benchmark}.
Estimation with the general-purpose optimizers almost always failed to converge towards parameter values with mean squared errors close to the (expected) variances underlying the simulations, indicating inefficient exploration of the parameter space (\emph{simulated annealing}) or convergence to local minima (\emph{quasi-Newton}). Since \emph{quasi-Newton} corresponds to step 3 of Algorithm~\ref{algo:nonpar}, the performance of \textit{ParDir-Nonlin} (steps 1 and 3) illustrates that the use of a parametric model to inform the initial guess of RKHS coefficients largely improved frequency of successful estimations (i.e., a mean squared error close to the nominal level), even though the parametric class qualitatively differed from the class used to generate the data (here, affine linear vs.\ saturable exponential). 
Refining the initial guess (step 1) for the quasi-Newton method further by iteratively solving the linearized RKHS model (step 2) allowed to considerably reduce runtime due to a closed-form solution of the linear inverse problem. 
Using the \emph{ParDir-AlyLin} variant allowed us to study whether the final quasi-Newton step in \emph{ParDir-AlyLin-Nonlin} was needed for the performance. 
As can be inferred, in some cases, the two steps \emph{ParDir-AlyLin} did not suffice to converge to nominal error levels, and hence the final quasi-Newton step proved necessary.
In all four scenarios (rich, sparse, noisy, multi), the benchmark test strongly supported the use of Algorithm~\ref{algo:nonpar} for use within the goodness-of-fit tests.

\subsection*{Goodness-of-fit testing for parametric maturation effect models}

For the goodness-of-fit testing problem, datasets were simulated for each of the four scenarios (rich, sparse, noisy, multi) using the saturable exponential model by \cite{robbie2012} as the true covariate-to-parameter relationship. 
Each of the three parametric classes (saturable exponential, affine linear, Michaelis-Menten) was considered as a null model, with the saturable exponential class representing a correctly specified model, and the other two models representing misspecified parametric models.
Subsequently, the proposed test statistics $T_1$ and $T_2$ were computed using Algorithms~\ref{algo:nonpar} and~\ref{algo:combined}, and their distribution under the parametric null hypothesis was approximated using $M=500$ Monte Carlo samples.
To determine the rejection frequency of the null hypothesis in each test, the entire procedure was repeated and averaged over 500 independent datasets.

The results of the goodness-of-fit tests are displayed in Tab.~\ref{tab:testresults} (see also  Section~\ref{sec:goftest-extra} for additional test statistics). 
Both tests approximately maintained the nominal Type~I error rate (here $\alpha=0.05$) in all four data scenarios. 
From theory, this behavior of the Monte Carlo approximation of the sampling distributions of the test statistics is expected whenever $\hat \tau$ is close to the unknown true parameter $\tau^*\in\mathcal{T}$.
The results thus indicated that the considered parametric model was estimated sufficiently well, even in a sparse data scenario.
Moreover, this result again highlighted the good performance of the numerical estimation algorithm.

As expected, the power to detect a misspecified parametric class proved to strongly depend on the considered data sampling scenario. 
For a rich data scenario, both test statistics had low Type~II error rates (0.4\%-2.4\%).
In the sparse data scenario as well as in the noisy data scenario, Type~II error rates increased considerably (54.2\%-82.3\%). 
In contrast, with a sampling scheme extending over several dosing intervals (scenario ``multi''), the goodness-of-fit test was again able to detect a model misspecification in presence of noisy data with Type~II error rates of 1.5\%-7.8\%.

We observed that the power depended on the test statistic (and on the parametric model class in case of the combined parametric/nonparametric alternative), though to a lesser degree than on the data scenario. 
Statistic $T_2$ resulted in lower Type~II error rates than $T_1$ for the affine linear class, and larger Type~II error rates for the Michaelis-Menten class.

In our setting, a purely parametric fit with the affine linear class resulted in lower mean squared errors compared to the Michaelis-Menten class, hence the affine linear model could be considered as a less severe model misspecification.
The results with $T_1$ compared to $T_2$ might indicate that the combined parametric/nonparametric formulation is beneficial for slight misspecifications, whereas the purely nonparametric formulation has advantages for more severe misspecifications.
Further investigations are warranted to elucidate the individual characteristics of the two test statistics. 
For the combined parametric/nonparametric alternative, the Type~II error shows an interesting parallel to the order of magnitude of cross-validated regularization parameters $\hat\lambda$. As the Type~II error, $\hat\lambda$ was lower for the affine linear model than for the Michaelis-Menten model. This seem to also indicate that to compensate for covariate model misspecification, a smaller nonparametric contribution is required  in the affine linear class than in the Michaelis-Menten class.

\begin{table}[htp!]
\noindent
\centering

\begin{tabular}{rr|rrrr}
& & \multicolumn{4}{c}{Type I error}\\
Model class of $H_0$& & rich & sparse & noisy & multi\\
\hline
\multirow{2}{*}{Saturable exponential} 
& $T_1$ &   5.2\% & 3.8\% & 5.2\% & 4.2\%\\
& $T_2$ &   5.2\% & 3.0\% & 5.6\% & 4.0\%\\
 \\
& & \multicolumn{4}{c}{Type II error}\\
Model class of $H_0$& & rich & sparse & noisy & multi\\
\hline
\multirow{2}{*}{Affine linear} 
& $T_1$ &   2.4\%  & 82.3\% & 83.8\% & 7.8\%\\
& $T_2$ &   0.6\%  & 66.2\% & 70.2\% & 2.4\%\\
\hline
\multirow{2}{*}{Michaelis-Menten} 
& $T_1$ &   0.4\% & 68.4\% & 54.2\% & 1.5\%\\
& $T_2$ &   1.0\% & 74.0\% & 68.8\% & 7.4\%\\
\end{tabular}

\caption{\label{tab:testresults} {\it 
Type~I and type~II errors in goodness-of-fit testing for the simulation study ``Effect of enzyme maturation on drug clearance'', for test statistics $T_1$ and $T_2$ (see Section S~\ref{sec:goftest-extra} for the complete set of test statistics, including the variants). 
For each of the simulation scenarios (rich, sparse, noisy, multi), 500 independent datasets were simulated with underlying saturable exponential model with parameters from Tab.~\ref{tab:parameters}. 
For each parametric hypothesis, the test statistics $T_1$ and $T_2$ were computed and their distribution under the null approximated with $M=500$ Monte Carlo samples. 
A level  $\alpha=0.05$ was taken as a decision threshold, i.e.~the empirical $0.05$-fractile of the Monte Carlo samples.
} }
\end{table}

\section*{Discussion}\label{sec:discussion}

We demonstrated the usefulness and practical applicability of the proposed goodness-of-fit tests for parametric covariate models in application to a relevant proof-of-concept study (``Effect of enzyme maturation on drug clearance'').
Due to the importance of covariate modeling in pharmacometric analyses, nonparametric goodness-of-fit tests have potential for a wide applicability.

Our simulation study was based on the meta-analysis of 22~separate studies in \cite{robbie2012}, which featured a very complex study design.
Although we made an effort to represent the essence of this meta-analysis, some specific aspects differed.
In \cite{robbie2012}, in addition to the covariates age and weight, the categorical covariates gender, ethnicity, and disease status were considered. The age distribution was non-uniform, with mainly young children and adults, since the disease is in young children (and adult were studied prior to children, as required by regulations).
Moreover, the sampling schedules for young children were sparser than for adults.
All of these effects could be integrated into a simulation study, but since they would make the presentation considerably more complex, we opted not to consider them here.
Moreover, the model in \cite{robbie2012} included random effects, which were not considered in our simulation study. 
Random effect models can be regarded as state-of-the-art in pharmacometric analyses \cite{standing2017}, and hence, this extension of the RKHS framework is of considerable importance. 
Therefore, this proof-of-concept study should be seen as the first step towards nonparametric goodness-of-fit testing in a non-linear mixed effects context. 
The consideration of random effects, however, require further extensions both on a theoretical and a numerical level, which were beyond the scope of the present manuscript.
 
In the main text, we compared the test statistics $T_1$ and $T_2$ defined on the observation space. In the supplement, we additionally considered a smoothed version $T_1^*$, as well as the corresponding statistics $S_1$, $S_1^*$ and $S_2$ on the parameter space (Section~\ref{sec:goftest-extra}).
The test statistics defined on the observation space consistently resulted in higher power compared to the respective test statistic defined on the parameter space (for all considered data scenarios and both misspecified parametric classes). 
On the observation space, the statistic $T_2$ based on a combined alternative showed superior power over the purely nonparametric statistics $T_1, T_1^*$ in the less regularized model.
The choice for a particular test statistic might thus be driven by the extent of regularization estimated by cross-validation.
The additional smoothing step differentiating $T_1^*$ from $T_1$ (and $S_1^*$ from $S_1$) was motivated by a theoretical analysis in the context of the direct problem, where (kernel density) smoothing was shown to increase robustness of test statistics \cite{haerdle1993}.
In our simulation study, such an improvement could be seen on the parameter space, $S_1^*$ showing superior power to $S_1$, but not on the observation space, where $T_1$ and $T_1^*$ resulted in almost identical test decisions.
In view of the required computational effort, $T_1$ might therefore be preferable to $T_1^*$, since the latter requires to solve an additional Tikhonov regularization problem during the smoothing step.
Here, further theoretical investigations are warranted.

The numerical algorithms presented have been carefully evaluated based on the simulation study.
Their robustness is achieved by approaching the nonlinear high-dimensional optimization problem stepwise through problems of increasing complexity.
Other heuristics to accomplish such ``coarse-graining'' could be envisaged, such as a first step of data pooling corresponding to individuals with similar covariates. The advantage of the proposed Algorithm~\ref{algo:nonpar}, however, is that is does not require such preprocessing steps. 

Unlike commonly used parametric models, our proposed kernel-based estimators allowed for very flexible covariate-to-parameter relationships, including non-monotonic functions.
In goodness-of-fit testing, the flexibility of nonparametric estimators is desirable since it allows to capture unexpected effects, e.g., potential nonmonotonic metabolic changes during puberty.
If a less variable nonparametric estimate of the covariate-to-parameter relationship were desired, hyperparameters such as the bandwidth of the Gaussian kernel could be adapted.
Indeed, it is known that optimal rates of testing and estimation may differ, with less regularization required for testing than for estimation \cite{ingster2003}.

For our proposed nonparametric goodness-of-fit test, we employed concepts from statistical learning, in particular kernel-based regularization techniques in the context of nonlinear statistical inverse problems with random design \cite{lu2013}.
The underlying RKHS framework is general and powerful, and hence the proposed setting  offers many possibilities for extension.
First, both the class of kernel functions and its hyperparameters (like the bandwidth of the Gaussian kernel) could be chosen to describe functions with a different degree of regularity, 
and a non-diagonal kernel structure could be exploited to simultaneously model parameters with a similar interpretation.
Also, regularization could be dealt with differently, for example through Landweber iterations rather than Tikhonov regularization \cite{hanke1995}.
The current approach uses a single regularization parameter $\lambda$. 
Inspired by \cite{horowitz2001}, however, one could use $\max_{\lambda\in \Lambda} T^{(\lambda)}$  over a suitably chosen grid $\Lambda$ of regularization parameters to propose an adaptive test (that is, a test for which no $\lambda$ has to be chosen); 
yet, a successful implementation of such an approach would require a normalization of the different test statistics $T^{(\lambda)}$, which seems difficult to achieve in the case of nonlinear inverse problems.
Finally, in the vector-valued RKHS setting, it is natural to generalize the regularization schemes by using different regularization parameters for different components of the RKHS function.\\

In summary, the flexibility of the RKHS framework renders the proposed goodness-of-fit tests very versatile;
our approach is envisioned to be beneficial for pharmacological applications that lack well-founded covariate models.


\newpage


\renewcommand{\theequation}{S\arabic{equation}}
\renewcommand\thesection{S\arabic{section}}
\setcounter{equation}{0}
\renewcommand\contentsname{Supplement}
\tableofcontents


\newpage

\section{Primal versus dual formulation of RKHS problems}
\label{sec:formulations}

A reproducing kernel Hilbert space (RKHS) is a particular function space, which can be uniquely characterized via a so-called \emph{kernel} $k$ (a positive definite and symmetric function).
A matrix-valued kernel function generates a \emph{vector-valued} RKHS, which can be used to model vector-valued functions such as the relationship between covariate(s) and several model parameters.

Here we summarize some basic definitions and results for $\mathbb{R}^p$-valued RKHS, more background can be found for instance in  
\cite{micchelli2005,caponnetto2007,caponnetto2008}. 
First, a function $k:\mathcal{X}\times \mathcal{X}\rightarrow \mathbb{R}^{p\times p}$ is called symmetric and positive definite if (i) $k(x,x')=k(x',x)^T$ for all $x,x'\in \mathcal{X}$ and (ii) $\sum_{i,j=1}^m \langle\alpha_i,k(x_i,x_j)\alpha_j\rangle_{\mathbb{R}^p}\geq 0$ for all $m\geq 1$, $\alpha_1,\dots,\alpha_m\in\R^p$ and $x_1,\dots,x_m\in\mathcal{X}$. 
Given such a kernel, there is a unique Hilbert space $\mathcal{H}$ of functions $h:\mathcal{X}\rightarrow \mathbb{R}^p$ such that 
$k_x:=k(\cdot,x)\in\mathcal{H}$ for all $x\in\mathcal{X}$,
and
\begin{equation}\label{eq:repr-prop}
\langle h(x),z\rangle_{\R^p} = \langle h,k_xz\rangle_{\mathcal{H}},\quad \text{for all } z\in\R^p, x\in\mathcal{X}, h\in\mathcal{H}.
\end{equation}
The space $\mathcal{H}$ is called the reproducing kernel Hilbert space associated with $k$, the function $k$ is often also called (reproducing) kernel, and \eqref{eq:repr-prop} is called reproducing property. 

The representer theorem (see \cite[Theorem~4.1]{micchelli2005}) guarantees the existence of $\hat h$ solving the Tikhonov regularization problem
\begin{equation}\label{eq:tikhonov-nonpar-supp}
\hat h := \argmin_{h\in\H}\left[\sum_{i=1}^n \Big\|y_i - G\big(h(x_i),x_i\big)\Big\|^2 + \lambda \big\|h\big\|^2_{\H}\right]
\end{equation}
within the finite-dimensional space
\begin{equation}\label{eq:shortdual-supp}
\left\{h_\alpha \in \H \,\Big|\,h_\alpha:=\sum_{i=1}^n k(\cdot,x_i)\alpha_i\; \text{ with } \; \alpha_1,\ldots,\alpha_n\in \R^p\right\}.
\end{equation}
For functions of the form \eqref{eq:shortdual-supp}, the RKHS norm $\|h\|_\H$ can be computed using the reproducing property \eqref{eq:repr-prop}, see next section.

\subsection{Dual formulation}
\label{sec:dual}

With the representer theorem, the solution $\hat h$ to \eqref{eq:tikhonov-nonpar-supp} can be sought within the set of candidate functions
\[\Big\{h_\alpha = \sum_{i=1}^n k(\cdot,x_i)\alpha_{i}:\quad \alpha_{1},...,\alpha_{n}\in\R^p\Big\}.\]
Since it will be of advantage later on, we consider the ordering
\[\alpha  = \begin{pmatrix}
\alpha_{\bigcdot\,1}\\
\vdots\\
\alpha_{\bigcdot\,p}
\end{pmatrix}\in\R^{np}, \quad\text{with}\quad \alpha_{\bigcdot\,l} := \begin{pmatrix}\alpha_{1,l}\\ \vdots\\ \alpha_{n,l}\end{pmatrix}\in\R^n.
\]
Accordingly, we define the kernel matrix 
\[\mathbb{K} := \begin{pmatrix}
\mathbb{K}_{11} & \cdots & \mathbb{K}_{1p}\\
\vdots & & \vdots \\
\mathbb{K}_{p1} & \cdots & \mathbb{K}_{pp}
\end{pmatrix}\in\R^{np\times np},\]
where
\[\mathbb{K}_{lm} = \begin{pmatrix} k_{lm}(x_1,x_1) & \ldots & k_{lm}(x_1,x_n)\\ \vdots & & \vdots\\ k_{lm}(x_n,x_1) & \ldots& k_{lm}(x_n,x_n)
\end{pmatrix}\in\R^{n\times n}.\]
With these expressions, we obtain the \emph{dual formulation} of the RKHS problem:
\begin{equation}
\argmin\limits_{\alpha\in\R^{np}}\left[\frac1n\sum_{i=1}^n\Big\|y_i - G\big(h_{\alpha}(x_i),x_i\big)\Big\|^2 + \lambda \big\|h_\alpha\big\|_{\H}^2\right]
\end{equation}
In this formulation, the RKHS norm can be computed as follows, using the reproducing property \eqref{eq:repr-prop}:
\[\big\|h_\alpha\big\|^2_{\mathcal{H}} = 
\sum_{i,j=1}^n \Big\langle k(\cdot,x_i)\alpha_{i}\,,\,k(\cdot,x_j)\alpha_{j} \Big\rangle_{\mathcal{H}} = 
\sum_{i,j=1}^n \alpha_j^Tk(x_j,x_i)\alpha_{i} = \alpha^T \mathbb{K} \alpha.
\]

\subsection{Primal formulation for scalar kernels}
\label{sec:primal}

Let us first consider a scalar kernel $k:\mathcal{X}\times\mathcal{X}\rightarrow\mathbb{R}$ (real-valued RKHS). The kernel $k$ has a \emph{finite-dimensional feature map representation}, if there is a mapping $\phi:\mathcal{X}\rightarrow \R^d$ such that
\[k(x,x') = \langle\phi(x),\phi(x')\rangle_{\mathbb{R}^{d}} \text{ for all } x,x'\in\mathcal{X}.\]
For example, polynomial kernels (i.e., $k(x,x') = (\langle x,x'\rangle + 1)^{m}$) admit a finite-dimensional feature map representation, but not a Gaussian kernel ($k(x,x') = \exp(\|x-x'\|^2/(2\sigma^2)$).

For a $d$-dimensional feature map representation with $d<n$, it is beneficial to cast the problem into the \emph{primal formulation}. To keep the notation simple, here and in the following we use the subscript to identify the type of formulation, with $\alpha,\hat\alpha$ for the dual formulation, $\beta,\hat\beta$ for the primal formulation,
$\gamma,\hat\gamma$ for the mixed formulation (next section). The primal formulation is derived as follows:
\[h_\alpha(x) = \sum_{i=1}^n k(x,x_i)\alpha_i = \Big\langle\underbrace{\sum_{i=1}^n\phi(x_i)\alpha_i}_{=:\beta\in\R^d},\phi(x)\Big\rangle_{\mathbb{R}^{d}} = \phi(x)^T\beta =: h_\beta(x),\]
reducing the $n$-dimensional problem of finding $\alpha = (\alpha_1,...,\alpha_n)$ to the $d$-dimensional problem of finding a solution in the space of functions
$\{h_\beta := \langle\beta,\phi(\cdot)\rangle_{\mathbb{R}^{d}} | \beta\in\R^d\}$,
which is, equipped with the norm $\|h\|_\H = \min\{ \|\beta\|_{\R^d}: h=h_{\beta},\beta\in\R^d\}$, the unique RKHS associated to $k$ (see e.g. \cite[Theorem~4.21]{steinwart2008}).
The resulting optimization problem in primal formulation is
\begin{equation}
\label{eq:primal}
\argmin\limits_{\beta\in\R^{d}}\left[\frac1n\sum_{i=1}^n\Big\|y_i - G\big(h_\beta(x_i),x_i\big)\Big\|^2 + \lambda \big\|h_\beta\big\|^2_{\mathcal{H}}\right].
\end{equation}
This optimization problem can be further simplified by replacing $\|h_\beta\|^2_{\mathcal{H}}$ by $\|\beta\|^2_{\mathbb{R}^{d}}$, since minimization of 
\begin{equation*}
\left[\frac1n\sum_{i=1}^n\Big\|y_i - G\big(h_\beta(x_i),x_i\big)\Big\|^2 + \lambda \big\|\beta\big\|^2\right].
\end{equation*}
 will automatically result in the minimum norm solution $\beta$ amongst all $\beta'$ such that $h_{\beta'}=h_\beta$. 

\subsection{Mixed formulation for diagonal kernels}
\label{sec:mixed}

Let us now consider a matrix-valued kernel function $k:\mathcal{X}\times\mathcal{X}\rightarrow\mathbb{R}^{p\times p}$ (vector-valued RKHS) with a diagonal kernel, i.e. such that 
\[k(x,x') = \begin{pmatrix}
k_1(x,x') & 0 & \cdots & 0\\
0 & k_2(x,x') & \ddots & \vdots\\ 
\vdots & \ddots & \ddots & 0\\
0 & \cdots & 0 & k_p(x,x')\\ 
\end{pmatrix}.\]
which means that the kernel matrix is block diagonal, $\mathbb{K} = \diag(\mathbb{K}_{11},...,\mathbb{K}_{pp})$, and that
\[h_\alpha(x) = \begin{pmatrix}
\sum_{i=1}^n k_1(x,x_i)\alpha_{i,1}\\
\vdots\\
\sum_{i=1}^n k_p(x,x_i)\alpha_{i,p}\\
\end{pmatrix}  =: \begin{pmatrix}
h_{\alpha_{\,\bigcdot\, 1},1}(x)\\
\vdots\\
h_{\alpha_{\,\bigcdot\, p},p}(x)\\
\end{pmatrix}. \]
Let $\{1,...,p\} = \mathcal{P} \cup \mathcal{D}$ be a decomposition such that 
\[l\in\mathcal{P}\quad\Rightarrow\quad k_l(x,x')=\langle\phi_l(x),\phi_l(x')\rangle_{\mathbb{R}^{d_l}}\]
with a $d_l$-dimensional feature map $\phi_l$ with $d_l < n$, i.e., for all indices within $\mathcal{P}$, the corresponding scalar kernel $k_l$ admits a lower-dimensional feature map representation and hence, it is beneficial to write it in the primal formulation. 
Analogous to the calculation in Sec.~\ref{sec:primal}, for any $l\in\mathcal{P}$, $h_{\alpha_{\,\bigcdot\, l},l}(x)$ can be rewritten as $h_{\beta_l,l}(x) := \langle\beta_l,\phi_l(x)\rangle_{\mathbb{R}^{d_l}}$ with $\beta_l\in\R^{d_l}$.

We now introduce a mixed primal-dual formulation for diagonal kernels. 
Define
\begin{itemize}
\item $\gamma = \begin{pmatrix}\gamma_1\\ \vdots \\ \gamma_p
\end{pmatrix}\in\R^d$ with $\gamma_l = \left\{\begin{tabular}{l c} $\alpha_{\,\bigcdot\, l}\in\R^n$ & if $l\in\mathcal{D}$\\
$\beta_l\in\R^{d_l}$ & if $l\in\mathcal{P}$\\
\end{tabular}\right.$ and $d := n|\mathcal{D}| + \sum\limits_{l\in P} d_l$,

\item $h_\gamma:=\begin{pmatrix}h_{\gamma_1,1}\\ \vdots \\ h_{\gamma_p,p}
\end{pmatrix}$ where $h_{\gamma_l,l}= \left\{\begin{tabular}{l c} $h_{\alpha_{\,\bigcdot\, l,l}}$ & if $l\in\mathcal{D}$\\
$h_{\beta_l,l}$ & if $l\in\mathcal{P}$\\
\end{tabular}\right.$.
\end{itemize}
The \emph{mixed (primal-dual) formulation} of Tikhonov regularization then reads

\begin{equation}\label{eq:mixed-tikhonov}
\argmin\limits_{\gamma\in\R^d} \left[\frac{1}{n}\sum_{i=1}^n\Big\| y_i - G\big(h_\gamma(x_i),x_i\big) \Big\|^2 + \lambda \big\|h_\gamma\big\|_\H^2\right].
\end{equation}
Similar to the primal problem \eqref{eq:primal}, the RKHS norm term can be simplified; to express it compactly, we define
\[\mathbb{D} := \begin{pmatrix} \mathbb{D}_1 & & 0\\ &\ddots& \\ 0 & & \mathbb{D}_p
\end{pmatrix}\in\R^{d\times d},\quad\text{where}\quad \mathbb{D}_l = \left\{\begin{tabular}{ll} 
$\mathbb{K}_{ll}$ & if $l\in\mathcal{D}$,\\
$I_{d_l}$ & if $l\in\mathcal{P}$,\\
\end{tabular}\right.
\]
which allows us to replace $\big\|h_\gamma\big\|_\H^2$ by $\gamma^T \mathbb{D} \gamma$.

\section{Solution of linear inverse problem}
\label{sec:lininvprob}

We now derive a closed-form expression for the solution of a linear inverse problem in mixed primal/dual form. As explained in the main text,
this expression is used in the numerical algorithms for a linearized version of the problem.

\subsection{Linearization}\label{sec:linearization}

We consider a linearization of the operator $Af(x)=G(f(x),x)$ at some function $f^*=g^*+h^*$ with $h^*\in\H$, which can be written as
\[G\big((g^*+h)(x),x\big) \approx G_\text{lin}\big((g^*+h)(x),x\big) := G\big(f^*(x),x\big) + D_\theta\big(f^*(x),x\big)\big(h(x)-h^*(x)\big).\]
Such a linearization is considered in step ``AlyLin'' of Algorithms~\ref{algo:nonpar} and~\ref{algo:combined} for $g^* = 0$ and $g^* = f_{\hat\tau}$, respectively, and $h^* = h^{(s-1)}_\text{lin}$, $s=1,...,\text{niter}$.
Using this notation, the resulting optimization problems can be written as
\begin{equation*}
\argmin\limits_{h\in\mathcal{H}}\left[\frac1n\sum_{i=1}^n\Big\|y_i - G_\text{lin}\big((g^*+h)(x_i),x_i\big)\Big\|^2 + \lambda \big\|h\big\|_{\H}^2\right].
\end{equation*}
Denoting 
\[L(x) := D_\theta\big(f^*(x),x\big)\] 
and
\[\ytrans_i := y_i - G\big(f^*(x_i),x_i\big) + L(x_i)h^*(x_i),\]
the problem can be written in the following way:
\begin{equation}
\argmin\limits_{h\in\mathcal{H}}\left[\frac1n\sum_{i=1}^n\Big\|\ytrans_i - L(x_i)\,h(x_i)\Big\|^2 + \lambda \big\|h\big\|_{\H}^2\right],
\end{equation}
which can be cast into a dual or, when appropriate, a primal or mixed formulation as described in Sec.~\ref{sec:formulations}. In the following, we derive the analytical solution of this problem in mixed formulation.

\subsection{Mixed formulation}\label{sec:lininv-mixed}

\paragraph{Mixed formulation in compact form.} 

To express the problem in mixed formulation in compact form, we introduce the following notation:
\begin{itemize}
\item The vectorized transformed data 
\[\ytrans := \begin{pmatrix}\ytrans_1\\ \vdots\\ \ytrans_n
\end{pmatrix}\in\R^{nq},\]

\item The vectorized parameters 
\[\theta  = \begin{pmatrix}
\theta_{\bigcdot\,1}\\
\vdots\\
\theta_{\bigcdot\,p}
\end{pmatrix}\in\R^{np}, \quad\text{with}\quad \theta_{\bigcdot\,l} := \begin{pmatrix}\theta_{1,l}\\ \vdots\\ \theta_{n,l}\end{pmatrix} = \begin{pmatrix} \Big(h_{\alpha}(x_1)\Big)_l\\ \vdots \\ \Big(h_{\alpha}(x_n)\Big)_l\end{pmatrix}\in\R^n,
\]
\item The feature matrices (for $l\in\mathcal{P}$):
\[\Phi_l := \begin{pmatrix}
\phi_l(x_1)^T \\ \vdots \\ \phi_l(x_n)^T
\end{pmatrix}\in\R^{n\times d_l},\]
\item The matrix
\[\mathbb{P} := \begin{pmatrix} \mathbb{P}_1 & & 0\\ &\ddots& \\ 0 & & \mathbb{P}_p
\end{pmatrix}\in\R^{np\times d},\quad\text{where}\quad \mathbb{P}_l = \left\{\begin{tabular}{ll} 
$I_n$ & if $l\in\mathcal{D}$,\\
$\Phi_l$ & if $l\in\mathcal{P}$,\\
\end{tabular}\right.
\]
which is defined such that $\theta = \mathbb{M}\gamma$ for $\mathbb{M}:= \mathbb{P}\,\mathbb{D}\in\R^{np\times d}$.

\item The linear forward operator 
\[\mathbb{L} = \begin{pmatrix}
\mathbb{L}_{11} & \cdots & \mathbb{L}_{1p}\\
\vdots & & \vdots \\
\mathbb{L}_{n1} & \cdots & \mathbb{L}_{np}
\end{pmatrix}\in\R^{nq\times np},\]
where $\mathbb{L}_{il}\in\R^{q\times n}$ is such that 
\[\big(\mathbb{L}_{il}\big)_{mj} = \left\{\begin{tabular}{ll} $L(x_i)_{ml}$ & if $i=j$,\\ 0 & otherwise.\end{tabular}\right.\]
This is simply the block-diagonal matrix $\diag\big(L(x_1),...,L(x_n)\big)$ with columns permuted such that per-parameter ordering of $\theta$ is matched.\\
\end{itemize}
Using this notation, we arrive at the following problem formulation:
\begin{equation}\label{eq:inverse-vv-mixed}
\argmin\limits_{\gamma\in\R^{d}} Q(\gamma), \text{ where } Q(\gamma) =  \big\|\ytrans- \mathbb{L}\,\mathbb{M}\,\gamma\big\|^2 + n\,\lambda\,  \gamma^T \mathbb{D}\gamma.
\end{equation}

\paragraph{Derivation of analytical solution.} 

Using the identities
\[ D_\gamma \Big(\big\|\ytrans- \mathbb{L}\,\mathbb{M}\,\gamma\big\|^2\Big) = 2(\mathbb{L}\,\mathbb{M}\,\gamma-\ytrans)^T\mathbb{L}\,\mathbb{M} \qquad \text{and}
\qquad D_\gamma \Big(\gamma^T\mathbb{D}\gamma\Big) = 2\gamma^T\mathbb{D},\]
we compute the derivative of $Q$:

\begin{align*}
DQ(\gamma) &= 2(\mathbb{L}\,\mathbb{M}\,\gamma-\ytrans)^T\mathbb{L}\,\mathbb{M} + 2n\lambda \gamma^T\mathbb{D}\\
 &= 2\gamma^T\mathbb{M}^T\mathbb{L}^T\mathbb{L}\,\mathbb{M} + 2n\lambda\gamma^T\mathbb{D} -2(\ytrans)^T \mathbb{L}\,\mathbb{M}\\ 
 &= 2\left[\big(\mathbb{M}^T\mathbb{L}^T\mathbb{L}\,\mathbb{M} + n\lambda\mathbb{D}\big)\gamma-\mathbb{M}^T\mathbb{L}^T\ytrans\right]^T\\ 
 &= 2\mathbb{D}\left[\big(\mathbb{P}^T\mathbb{L}^T\mathbb{L}\,\mathbb{M} + n\lambda I_{d}\big)\gamma-\mathbb{P}^T\mathbb{L}^T\ytrans\right]^T\\ 
\end{align*}
Therefore, $DQ(\gamma)=0$ if $\big(\mathbb{P}^T\mathbb{L}^T\mathbb{L}\,\mathbb{M} + n\lambda I_d\big)\gamma=\mathbb{P}^T\mathbb{L}^T\ytrans$, and
\[\hat\gamma := \big(\mathbb{P}^T\mathbb{L}^T\mathbb{L}\,\mathbb{M} + n\lambda I_d\big)^{-1}\mathbb{P}^T\mathbb{L}^T\ytrans\]
is a minimizer of \eqref{eq:inverse-vv-mixed}.

\paragraph{Special case.}

In the case of a direct problem, $q=p$ and $\mathbb{L} = I_{np}$. Furthermore, in dual formulation $\mathbb{M} = \mathbb{D} = \mathbb{K}$ and $\mathbb{P} = I_{np}$, which yields the well-known formula (see e.g. \cite[Eq.~5]{alvarez2012})
\[\hat\alpha := \big(\mathbb{K} + n\lambda I_{np}\big)^{-1}\ytrans.\]

\clearpage
\section{Miscellaneous}\label{sec:misc}

\subsection{Analytical solution of the ODE system \eqref{eq:ODE-1}--\eqref{eq:ODE-2}}\label{sec:Gmaturation}

Since the ODE system \eqref{eq:ODE-1}--\eqref{eq:ODE-2} is linear, it can be solved analytically. Setting
\[\zeta_{1/2} := -\frac{1}{2}\left(\frac{\text{CL}+Q}{V_1}+\frac{Q}{V_2} \mp \sqrt{\left(\frac{\text{CL}+Q}{V_1}+\frac{Q}{V_2}\right)^2 - 4 \frac{\text{CL}\,Q}{V_1\,V_2}}\right)\]
(the eigenvalues of the ODE right-hand side),
the analytical expression for $C_1$ with initial condition $C_1(0)=\frac{\text{D}}{V_1}$ (i.e., after an absolute dose D) is given by
\[C_1(t) = \frac{\text{D}}{V_1} \frac{1}{\zeta_1 - \zeta_2} \left[ \left(\zeta_1+\frac{Q}{V_2}\right)e^{\zeta_1\, t} - \left(\zeta_2+\frac{Q}{V_2}\right)e^{\zeta_2\, t}\right],\]
see \cite{dargenio2019,koch2012} for reference.

\subsection{Simulated data for the different scenarios}\label{sec:sim-data}

\begin{figure}[htp!]
\centering
\includegraphics[width=\textwidth]{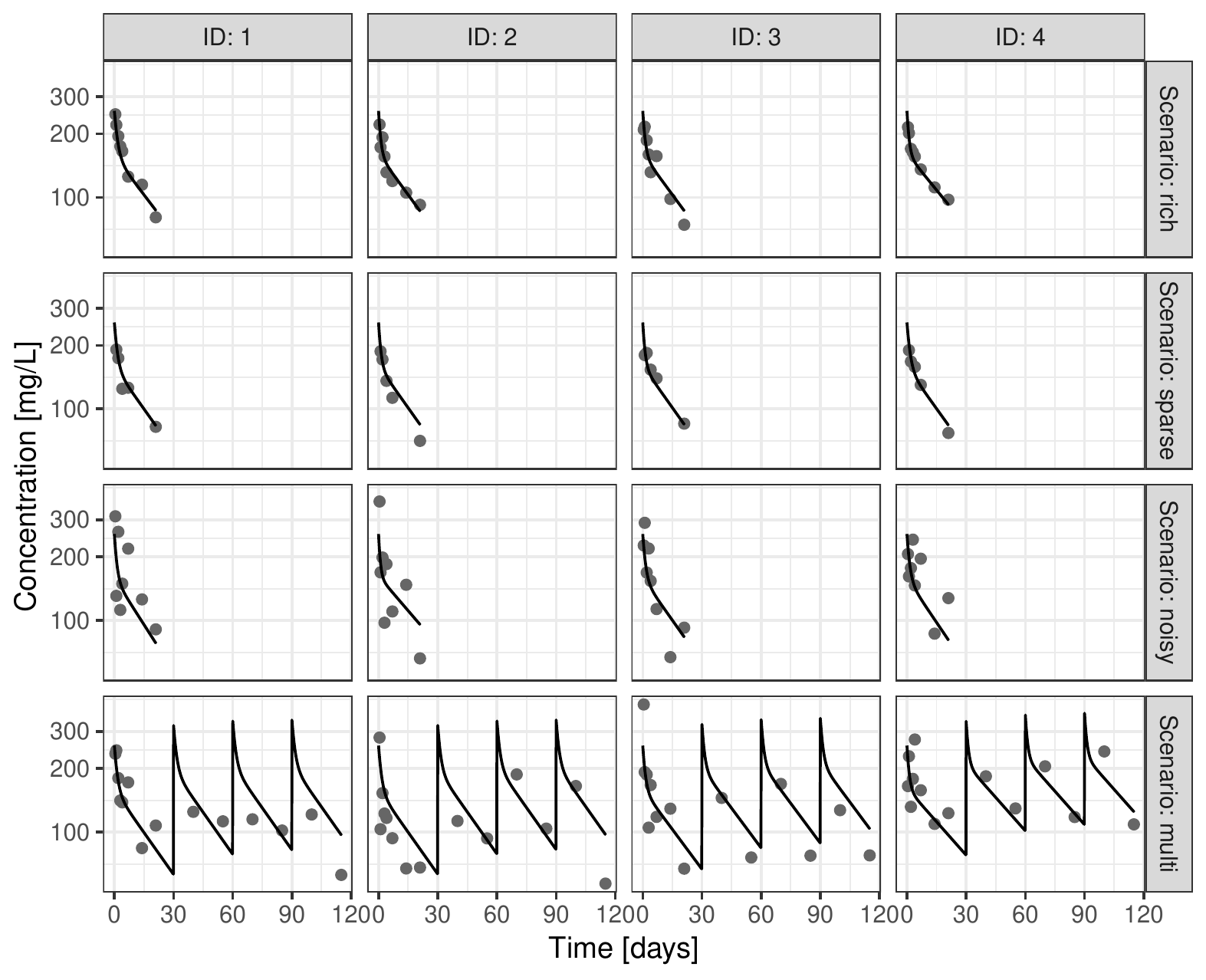}
\caption{\label{fig:conc} {\it
Simulated data and prediction with data-generating mechanistic/covariate model. For each scenario (rich, sparse, noisy, and multi), simulated data (grey dots) and model predictions (black lines) are shown exemplarily for four individuals.}}
\end{figure}

\clearpage

\subsection{Pseudocode for combined parametric/RKHS algorithm}\label{sec:app-algo2}

Since using the parametric part appears explicitly in the problem formulation, the parametric estimate $f_{\hat\tau}$ is used in a different way. 
The parametric part determined in step 1 is fixed in the optimization problem, and steps 2 and 3 are used to determine the nonparametric part.
As an initial guess for the nonparametric part in step 2, the zero function can be used, and hence, no further direct problem needs to be solved in step 1 to determine appropriate initial conditions.
Conceptually, the solution of \eqref{eq:tikhonov-combined} over $\mathcal{T}\times \H$ is split into the parametric and nonparametric part.
Although the parametric part of \eqref{eq:tikhonov-combined} may be different to the least squares estimate $\hat \tau$ (used in step 1), the benchmark in Sec.~\ref{sec:benchmark-combined} showed a good agreement of these quantities, and no relevant improvement by solving \eqref{eq:tikhonov-combined} over $\mathcal{T}\times \H$ jointly.

\setcounter{algocf}{1}

\begin{algorithm}[htp!]
\DontPrintSemicolon
\SetKwFunction{LevenbergMarquardt}{LevenbergMarquardt}
\SetKwFunction{LinearizeModel}{LinearizeModel}
\SetKwFunction{QuasiNewton}{QuasiNewton}
\SetKwData{niter}{niter}
\BlankLine
\BlankLine
\tcp{Step 1:~"Par"}
$\text{res} := \left[\tau \,\mapsto\, \Big(G(f_\tau(x_1),x_1)-y_1,...,G(f_\tau(x_n),x_n)-y_n\Big)\right]$\;
$\hat\tau \longleftarrow \text{minimize } \|\text{res}(\tau)\|^2 \text{ using Levenberg-Marquart with initial guess } \tau_0$\;
\BlankLine
\BlankLine
\tcp{Step 2:~"AlyLin"}
$h_\text{lin}^{(0)} \longleftarrow 0$\;
\For{$s =  1:\niter$}{
$G^{(s)}_\text{lin} \longleftarrow \text{linearize model } G \text{ at function } f_{\hat\tau}+ h_\text{lin}^{(s-1)}$\;
$Q^{(s)}_\text{lin} := \left[h \mapsto \sum\limits_{i=1}^n \Big\|y_i-G^{(s)}_\text{lin}\big(f_{\hat\tau}(x_i)+h(x_i),x_i\big)\Big\|^2 + \lambda \big\|h\big\|^2_\H\right]$\;

$h_\text{lin}^{(s)} \longleftarrow \text{ minimize } Q^{(s)}_\text{lin}(h) \text{ over } \H \text{ analytically }$\;
}
\BlankLine
\BlankLine
\BlankLine
\tcp{Step 3:~"Nonlin"}
$Q := \left[h \mapsto \sum\limits_{i=1}^n \Big\|y_i-G\big(f_{\hat\tau}(x_i)+h(x_i),x_i\big)\Big\|^2 + \lambda \big\|h\big\|^2_\H\right]$\;
$h_\text{nonlin} \longleftarrow \text{minimize } Q(h) \text{ using quasi-Newton with initial guess } h_\text{lin}^{(\niter)}$\;
\BlankLine
\BlankLine
\BlankLine
\tcp{Output}
$\tilde{f}^{(\lambda)} \longleftarrow f_{\hat\tau} + h_\text{nonlin}$\;
\KwRet{$\tilde{f}^{(\lambda)}$}
\caption{\label{algo:combined}
Par-AlyLin-Nonlin for problem \eqref{eq:tikhonov-combined}.
}
\end{algorithm}

\clearpage

\subsection{Benchmark for combined parametric/RKHS algorithm}\label{sec:benchmark-combined}

\begin{figure}[htp!]
\centering
\includegraphics[width=\textwidth]{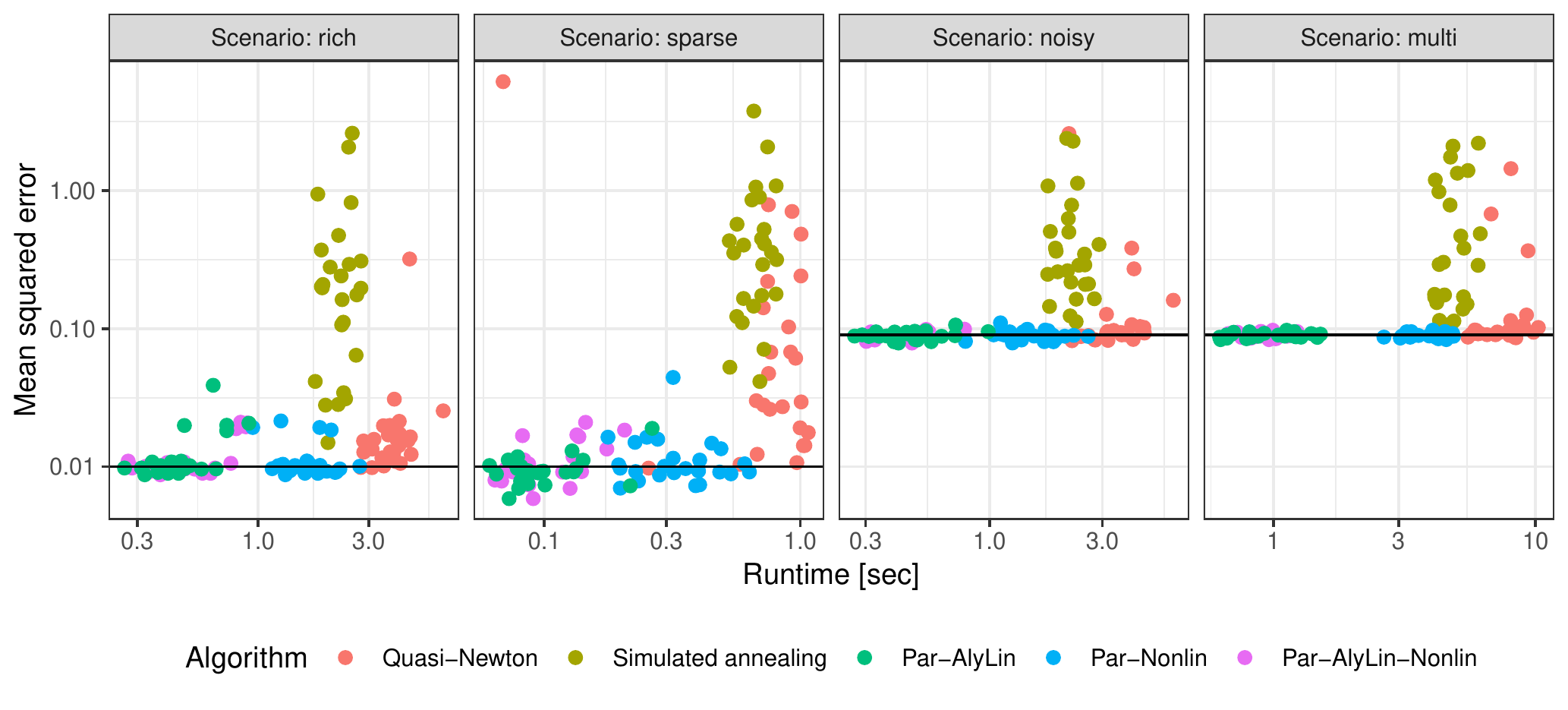}
\caption{\label{fig:benchmark2} {\it 
Benchmark of algorithms for solving the Tikhonov-type regularization problem \eqref{eq:tikhonov-combined} in the simulation study ``Effect of enzyme maturation on drug clearance''. 
We simulated 25 independent datasets for each of the four considered data scenarios (rich, sparse, noisy, multi) and benchmarked the following estimation algorithms: a quasi-Newton method, simulated annealing, the proposed Algorithm~\ref{algo:combined} (steps 1-3); and two variants of it:  Par-AlyLin (only steps 1 and 2) and Par-Nonlin (only steps 1 and 3).
Each dot represents runtime and mean squared error for one of the 25 simulated datasets.
The black horizontal line indicates the error variance $\sigma^2$ used during the simulation.
}}
\end{figure}

In addition to the benchmark of algorithms for the nonparametric problem \eqref{eq:tikhonov-nonpar}, we also compared algorithms for solving the combined parametric/nonparametric problem \eqref{eq:tikhonov-combined}.
The results of this benchmark are displayed in Fig.~\ref{fig:benchmark2}. 
Similar results as in the nonparametric case were obtained: 
estimation with the general-purpose optimizers almost always failed to converge towards nominal error levels; 
the use of a parametric model to inform the initial guess of RKHS coefficients largely improved frequency of successful estimations; 
and the AlyLin step considerably reduced runtime. 
In contrast to the purely nonparametric case, the final quasi-Newton step in Par-AlyLin-Nonlin offered only a minor improvement over Par-AlyLin.
Again,  the benchmark test strongly supported the use of Algorithm~\ref{algo:combined} for use within the goodness-of-fit tests in all four scenarios (rich, sparse, noisy, multi).

\clearpage

\subsection{Goodness-of-fit testing for parametric maturation effect models: results for additional test statistics}\label{sec:goftest-extra}

To address the potential difference in smoothness between the parametric and nonparametric estimator, we additionally consider a smoothed parametric estimator
\begin{equation}\label{eq:tikhonov-smoothed}
\hat f_{\hat\tau}^{(\lambda)} := \argmin_{h\in\H}\left[\sum_{i=1}^n \Big\|\tilde y_i - G\big(h(x_i),x_i\big)\Big\|^2 + \lambda \big\|h\big\|^2_{\H}\right],\end{equation}
using artificial data $\tilde y_i=G\big(f_{\hat \tau}(x_i),x_i\big)$ based on the parametric estimator $f_{\hat \tau}(x_i)$, rather than the observed data $y_i$.
Based on this, we consider the test statistic
\begin{equation}\label{eq:T2}
T_1^* := \sum_{i=1}^n \Big\|G\Big(\hat f_{\hat \tau}^{(\lambda)}(x_i),x_i\Big) - G\Big(\hat f^{(\lambda)}(x_i),x_i\Big)\Big\|^2.
\end{equation}
Whereas $T_1$ compares the predictions with a parametric estimator vs.~predictions of an RKHS estimator, $T_1^*$ additionally enforces the same smoothness of parametric and RKHS estimators through the additional Tikhonov regularization step. 

Algorithm~\ref{algo:nonpar} can also be employed to obtain the smoothed parametric estimate 
$\hat f_{\hat\tau}^{(\lambda)}$ solving \eqref{eq:tikhonov-smoothed}.
Using the parametric model $f_{\hat\tau}$ to be smoothed in step~1 of the algorithm, a very good initial estimate for $\hat f_{\hat\tau}^{(\lambda)}$ is obtained, rendering this problem computationally simpler than \eqref{eq:tikhonov-nonpar}, where the parametric class might be misspecified.

In the combined parametric/RKHS formulation, no distinction between non-smoothed and smoothed classes is required;
the additional regularization step that differentiates $T_1$ from $T_1^*$ in the nonparametric case has no effect in the combined parametric/RKHS case:
the estimator $\tilde f^{(\lambda)}$ based on the artificial data $\tilde{y}_i = G\Big(f_{\hat \tau}(x_i),x_i\Big)$ instead of $y_i$ coincides with $f_{\hat \tau}$ since the functional in \eqref{eq:tikhonov-combined} is 0 in this function; any RKHS contribution would be penalized. Therefore, $T_2^*=T_2$.

Analogously to the statistics $T_1$, $T_1^*$ and $T_2$ defined on the observable space, we define statistics $S_1$, $S_1^*$, $S_2$ defined on the parameter space, more precisely, 
\begin{align*}
S_1 &:= \sum_{i=1}^n \Big\|f_{\hat \tau}(x_i) - \hat f^{(\lambda)}(x_i)\Big\|^2;\\
S_1^* &:= \sum_{i=1}^n \Big\|\hat f_{\hat \tau}^{(\lambda)}(x_i) - \hat f^{(\lambda)}(x_i)\Big\|^2;\\
S_2 &:= \sum_{i=1}^n \Big\|f_{\hat \tau}(x_i) - \tilde f^{(\lambda)}(x_i)\Big\|^2.
\end{align*}
The results of the goodness-of-fit tests for the simulation study ``Effect of enzyme maturation on drug clearance'' corresponding to all considered test statistics ($T_1$, $T_1^*$, $T_2$, $S_1$, $S_1^*$, $S_2$) are displayed in Tab.~\ref{tab:testresults2}. 
All test statistics maintained nominal Type~I error levels, but the power of $S_j$ under a misspecified parametric class was lower than the power of $T_j$ ($j=1,2,3$), for all four considered data scenarios (rich, sparse, noisy, multi) and both types of misspecified models (affine linear, Michaelis-Menten). 

\begin{table}[htp!]
\noindent
\centering

\begin{tabular}{rr|rrrr}
& & \multicolumn{4}{c}{Type I error}\\
Model class of $H_0$& & rich & sparse & noisy & multi\\
\hline
\multirow{6}{*}{Saturable exponential} 
& $T_1$ &   5.2\% & 3.8\% & 5.2\% & 4.2\%\\
& $T_1^*$ & 5.2\% & 4.2\% & 5.2\% & 4.0\%\\
& $T_2$ &   5.2\% & 3.0\% & 5.6\% & 4.0\%\\
& $S_1$ &   1.6\% & 0.4\% & 0.6\% & 2.0\%\\
& $S_1^*$ & 5.0\% & 1.8\% & 3.2\% & 4.4\%\\
& $S_2$ &   5.0\% & 3.4\% & 6.0\% & 5.0\%\\
 \\
& & \multicolumn{4}{c}{Type II error}\\
Model class of $H_0$& & rich & sparse & noisy & multi\\
\hline
\multirow{6}{*}{Affine linear} 
& $T_1$ &   2.4\%  & 82.3\% & 83.8\% & 7.8\%\\
& $T_1^*$ & 2.6\%  & 82.3\% & 84.2\% & 7.8\%\\
& $T_2$ &   0.6\%  & 66.2\% & 70.2\% & 2.4\%\\
& $S_1$ &   44.0\% & 99.4\% & 99.0\% & 68.6\%\\
& $S_1^*$ & 16.2\% & 96.6\% & 85.8\% & 26.6\%\\
& $S_2$ &   0.8\%  & 68.8\% & 74.2\% & 3.6\%\\
\hline
\multirow{6}{*}{Michaelis-Menten} 
& $T_1$ &   0.4\% & 68.4\% & 54.2\% & 1.5\%\\
& $T_1^*$ & 0.6\% & 68.7\% & 57.8\% & 1.5\%\\
& $T_2$ &   1.0\% & 74.0\% & 68.8\% & 7.4\%\\
& $S_1$ &   0.2\% & 94.4\% & 64.8\% & 0.9\%\\
& $S_1^*$ & 0.0\% & 77.7\% & 47.2\% & 0.2\%\\
& $S_2$ &   2.2\% & 82.0\% & 82.4\% & 14.0\%\\
\end{tabular}

\caption{\label{tab:testresults2} {\it 
Type~I and type~II errors for all considered test statistics in goodness-of-fit testing for the simulation study ``Effect of enzyme maturation on drug clearance''. 
For each of the data scenarios (rich, sparse, noisy, multi), 500 independent datasets were simulated with underlying saturable exponential model with parameters from Tab.~\ref{tab:parameters}. 
For each parametric hypothesis, the test statistics $T_1$, $T_1^*$, $T_2$, $S_1$, $S_1^*$ and $S_2$ were computed and their distribution under the null approximated with $M=500$ Monte Carlo samples. 
A level  $\alpha=0.05$ was taken as a decision threshold, i.e.~the empirical $0.05$-fractile of the Monte Carlo samples.
} }
\end{table}


\end{document}